\documentclass[aps,prstab,amsmath,twocolumn]{revtex4-2}

\usepackage{siunitx,graphicx,amsfonts}

\begin{document}

\title{Automatic adjustment of undulator optics for FELs}

\author{Z. Chen}
\email{chenzc@sari.ac.cn}
\affiliation{Shanghai Advanced Research Institute}

\author{B. Faatz}
\affiliation{Shanghai Advanced Research Institute}

\begin{abstract}
In this paper, we describe a way to automatically adjust the quadrupole focusing along the undulator to avoid the instabilities, taking into account energy change and undulator focusing. The procedure is more generalized and applicable to any strongly focusing (planar) undulator.
\end{abstract}
\maketitle

\section{Introduction}

For machines like SHINE \cite{FEL2017}, with high beam energy, the undulator focusing has little influence on the beam size. Even if the beam energy is reduced from the nominal value of 8 GeV to 5 GeV or 2.5 GeV, the statement is true. This situation is different for machines like SXFEL~\cite{Zhao2017}, operated at lower energy. Here, even with a much lower K-parameter of the undulator, the undulator focusing can play a dominant role.
\begin{figure}[!t]
	\centering
	 \includegraphics*[width=\columnwidth]{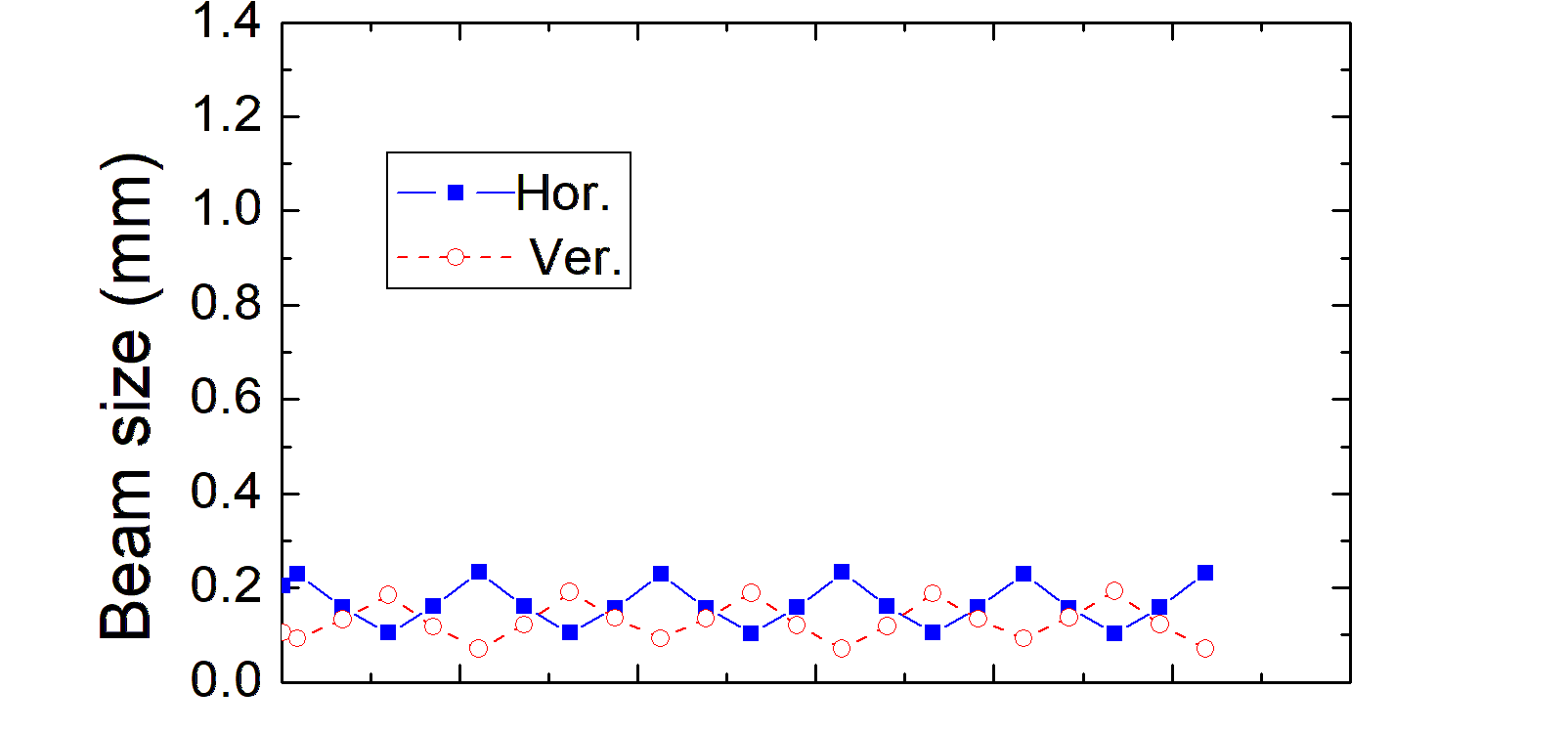}
	 \includegraphics*[width=\columnwidth]{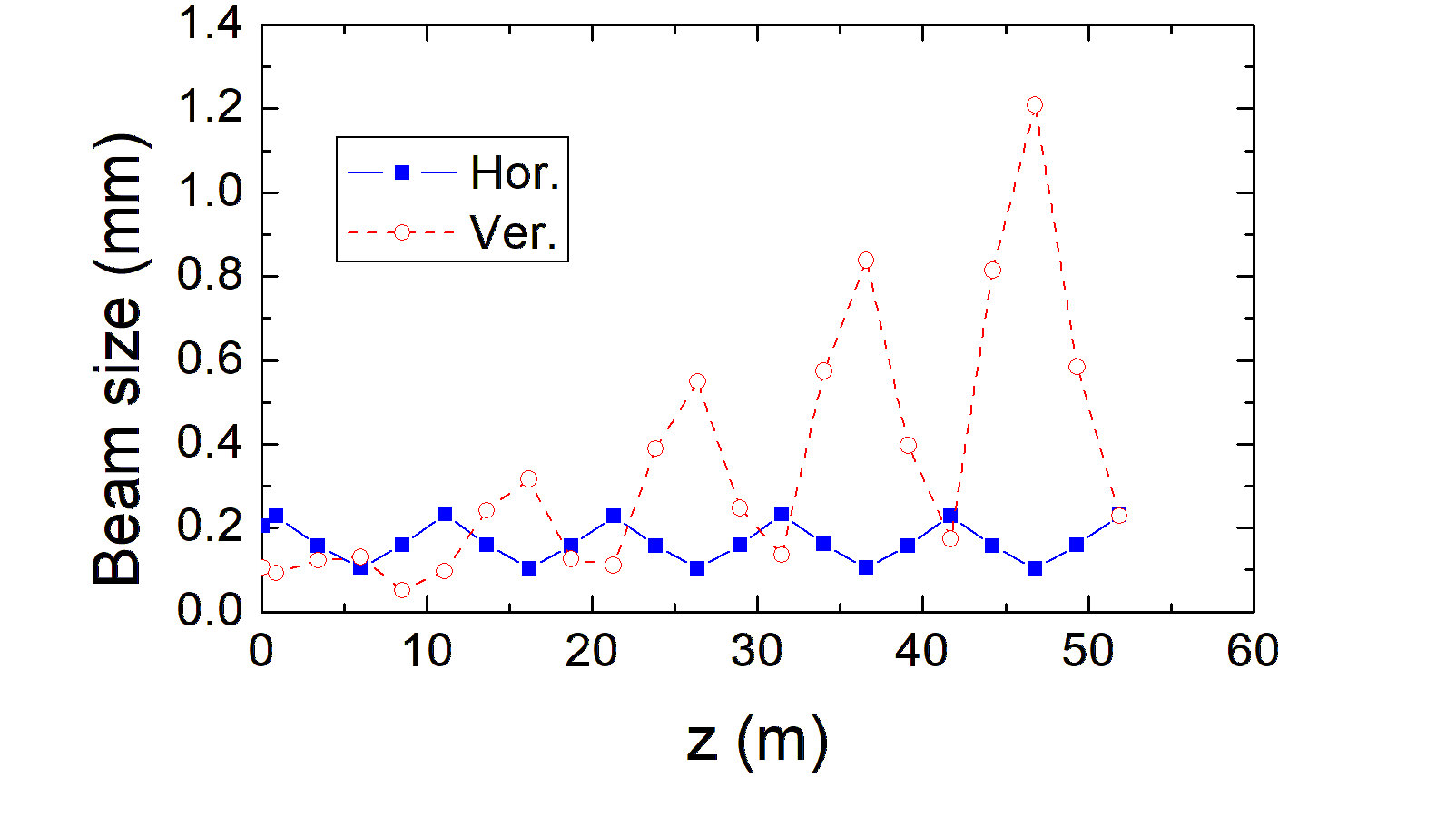}
	 	\caption{Beamsize along the undulator for standard quadrupole settings with undulators opened (top) and undulators closed (bottom) using parameters of the SBP undulator at SXFEL in Shanghai \cite{Zhao2017} for a beam energy of 1 GeV. As can be seen, the beam size increases along the undulator by an order of magnitude compared to the case with all undulators opened.}
	\label{fig:SXFEL-instability}
\end{figure}
Because of the undulator focusing, which increases with decreasing undulator gap, the focusing system can become unstable, as can be seen in Fig.~\ref{fig:SXFEL-instability} for parameters of the SBP-undulator at  SXFEL. The top figure shows the situation when the undulators are open, the bottom figure when they are closed, both for a beam energy of 1~GeV, which is well within the range of energies foreseen for standard operation. Theoretically, the effect that causes the instability has been very well described in ref.~\cite{Quattromini2012}, using as an example the SPARC FEL parameters. The same effect was expected in advance at FLASH2 at a very low energy, which is why the focusing in that FEL is adjusted automatically by the undulator server, based on the undulator gap and beam energy \cite{Faatz2017}. The wavelength range at which the FLASH2 beamline can produce radiation was extended because of the increased stability of the focusing at low beam energy by almost a factor of two, without touching the focusing manually. The server adjusts all quads including the one directly in front of the first undulator and the quad directly behind the last undulator. Quadrupoles further upstream and behind the undulator were not adjusted at all. This means, that the matching of the electron beam into the undulator was not changed. In case this is needed to optimize the FEL performance, a separate procedure has to be used. However, this will not change the stability of the solution inside the undulator itself and is not considered in this paper.

The reason for the instabilities can be understood as follows. Starting with a normal FODO-structure with all undulators open, the beam is focused and stable in both planes by having an alternating focusing and defocusing quad along the undulator. As the undulators are closed, effectively an additional focusing element is added to the vertical plane and the focal point of quad and undulator combined moves for a specific undulator gap in front of the vertically defocusing quad, which makes the system unstable. The instability can be avoided by changing the system from a conventional FODO to a FOFO structure, where all quadrupoles defocus vertically, therefore moving the focal point further downstream and making the system stable again. With both quads now focusing horizontally, the horizontal beam size will decrease. But with further increased undulator focusing at even smaller gaps, at some point the focal point of the undulator alone is short enough to move its focal point in front of the defocusing quad, making the system again unstable. This instability can be again avoided by changing back from FOFO to FODO, but in this case with a very short focal point. In order to avoid an instability in the horizontal plane, the horizontally focusing quad needs to have its focal point a fraction beyond the defocusing, strong quad. This causes extremely large variation of the beta-function, which is obviously not desirable, but the only way to obtain a stable solution. This way, what is referred to forbidden zones in \cite{Quattromini2012} can be avoided, but only for a linear undulator, as will be discussed in the last section.

In this paper, we will start with a theoretical model to describe the system. With this model, three cases will be studies, namely a very short (``zero-length'') undulator, an undulator filling the complete space between the quadrupoles and the general case.  Finally, some general conclusion and remaining challenges will be discussed.

\section{Model used}

\begin{figure*}[!t]
	\centering
	 \includegraphics*[width=0.9\textwidth]{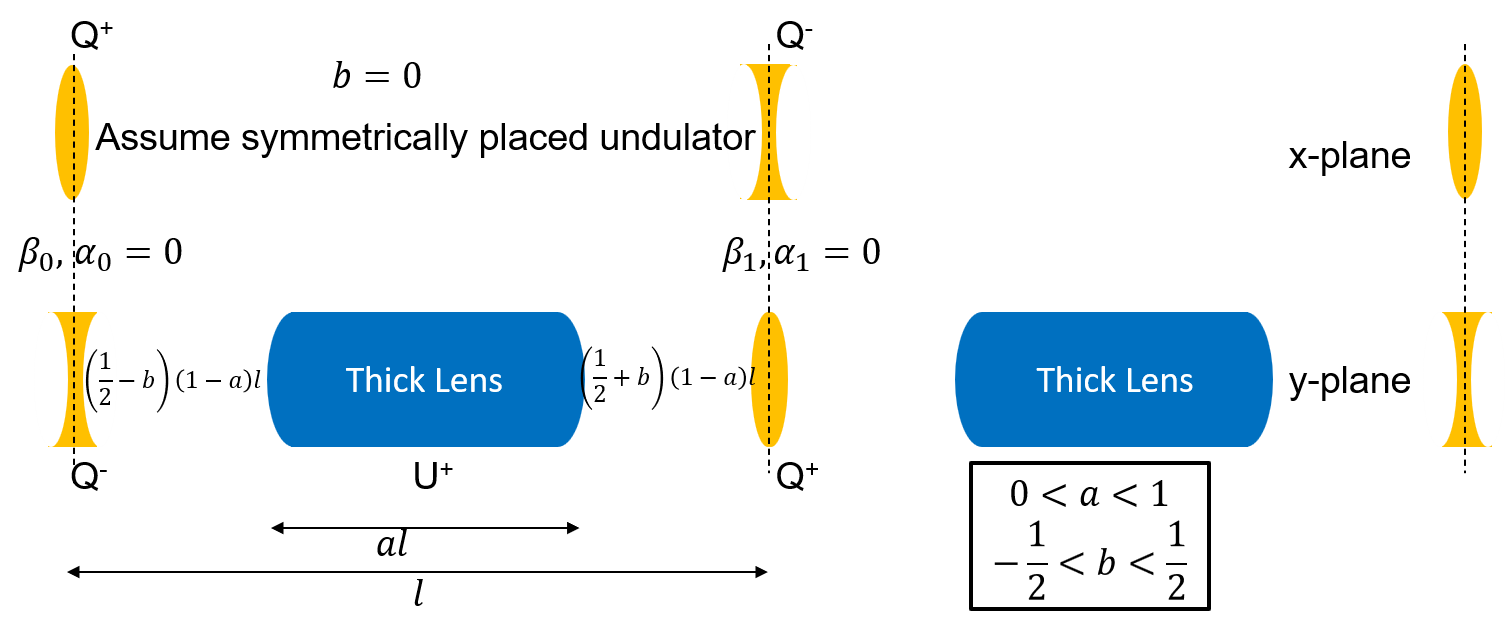}

	\caption{Geometry of the structure under study. In the horizontal plane, the structure consists of quadrupoles, in the vertical plane, there is an undulator placed between the quads, adding a focusing term with its strength depending on the undulator gap.}
	\label{fig:Geometry}
\end{figure*}

We use the notation shown in Fig.~\ref{fig:Geometry}. Quads are denoted by $Q_1$ and $Q_2$ with their strength given as inverse focal length normalized to the distance between the quad centers $q_1/l$ and $q_2/l$, where $l$ is the distance between the quads. The undulator parameters is given by $\sqrt{a}\xi=K_{RMS}k_u a\, l/\gamma$, where $K_{RMS}$ is the RMS-value of the undulator strength, $\lambda_u=2\pi/k_u$ is the undulator period, $\gamma$ is the Lorentz factor and $l_u=a\, l$ is the undulator length with $0<a<1$. Between quads and undulator are drifts with length of $(1/2\pm b)(1-a)l$, where $-1/2<b<1/2$ and $b=0$ means that the undulator is centered. For symmetry reasons, calculations start in the center of a quad. Calculating from quad center to quad center, there is a quad, drift, undulator, drift and quad. The matrix $M(q_1,q_2,a,b,\xi, l )$ of this system, which is half a FODO-period, is
 \begin{widetext}
 \begin{equation}
 \begin{pmatrix}
M_{11}				 &M_{12} \\
M_{21}			& M_{22}		 \\
\end{pmatrix}
=
\begin{pmatrix}
B_{a,b}(\xi)-A_{a,b}(\xi)q_2				 &A_{a,b}(\xi) l  \\
M_{21}				& B_{a,-b}(\xi)-A_{a,b}(\xi)q_1		 \\
\end{pmatrix}
\end{equation}
\end{widetext}
where the functions $A_{a,b}(\xi)$ and $B_{a,\pm b}(\xi)$ will be discussed in the appendix. Note that $M_{1,1}$ and and $M_{2,2}$ have a different sign in $b$, which is due to the fact that the drift in front and behind the undulator have opposite sign in $b$. $M_{21}$ can be determined easily by using that the matrix has a determinant of 1. The total system, containing a full FODO-period, can be written as $M(q_1,q_2,a,b,\xi, l )\cdot M(q_2,q_1,a,b,\xi, l )$. Since we are looking for a periodic solution of this system, we need to find its eigenvalues and eigenvectors. Because they have to be real and the system is symplectic, these solutions can only be found if $-2<{\rm Trace} (M(q_1,q_2,a,b,\xi, l )\cdot M(q_2,q_1,a,b,\xi, l ))<2$, where 2 is the dimension of the matrix. Therefore, instead of deriving the complete matrix, we are only interested the stability function
$S_{a,b}(q_1,q_2,\xi)$, which is derived in the appendix and results in
 \begin{equation}
0\le \left(B_{a,0}(\xi)+A_{a,b}(\xi)q_1\right)\left(B_{a,0}(\xi)+A_{a,b}(\xi)q_2\right)\le 1\,.
\label{Eq:stability2}
\end{equation}
Note that only $B_{a,0}(\xi)$ shows up in this equation, and only $B_{a,b}(\xi)$ depends on the location of the undulator.

A number of properties of Eq.~\eqref{Eq:stability2} are clear immediately. They are discussion in further detail in the appendix.
\begin{itemize}
\item When $A_{a,b}(\xi)=0$, the quads have no influence on the stability of the system, and the stability reduces to $B_{a,0}(\xi)^2\le 1$. It is shown in the appendix that $B_{a,b}(\xi)=B_{a,-b}(\xi)=\pm 1$ for $b=0$ and larger otherwise. Therefore, at these points the system is only stable when the undulator is placed centered between the quads.
\item When $B_{a,0}(\xi)=0$, the stability becomes $0<A_{a,b}(\xi)^2q_1q_2<1$, which means that the quads need to have equal sign, more specifically, they need to be horizontally focusing and change from FODO to FOFO lattice.  No general statement about the stability can be made.
\end{itemize}

\section{Analysis of different systems}

In the following sections, we study a normal FODO lattice, a thin planar undulator which is focusing vertically, a planar undulator filling the complete  space between the quads and a planar undulator filling only part of the space.

\subsection{FODO lattice}

\begin{figure}[!t]
	\centering
	\includegraphics*[width=0.7\columnwidth]{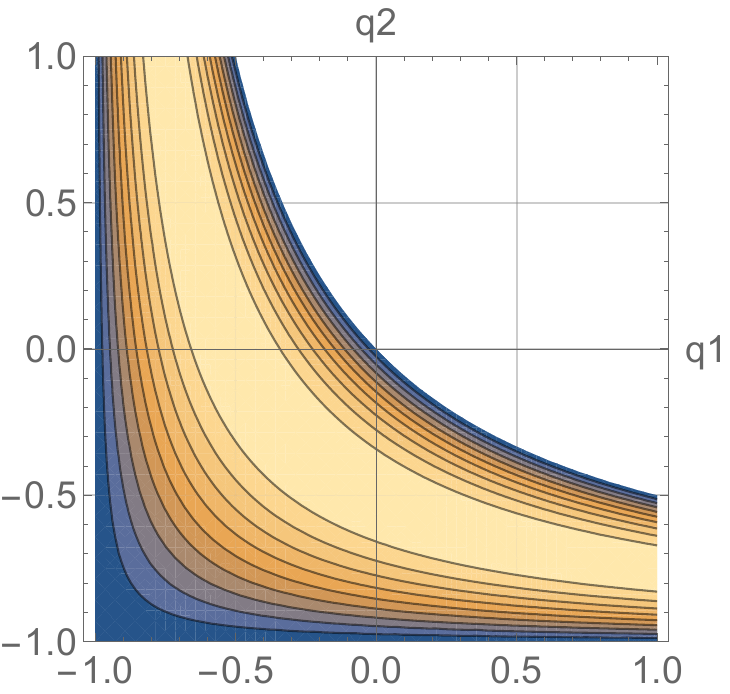}
	\includegraphics*[width=0.07\columnwidth]{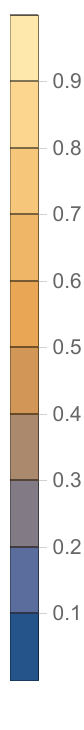}
 	\includegraphics*[width=0.7\columnwidth]{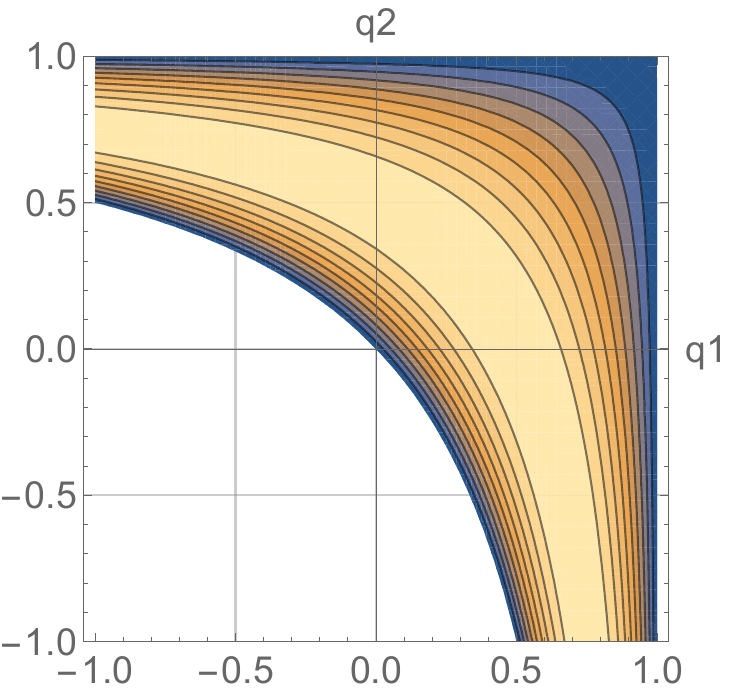}
 	\includegraphics*[width=0.07\columnwidth]{Stability-FODO-Barlegend.pdf}
 	\includegraphics*[width=0.7\columnwidth]{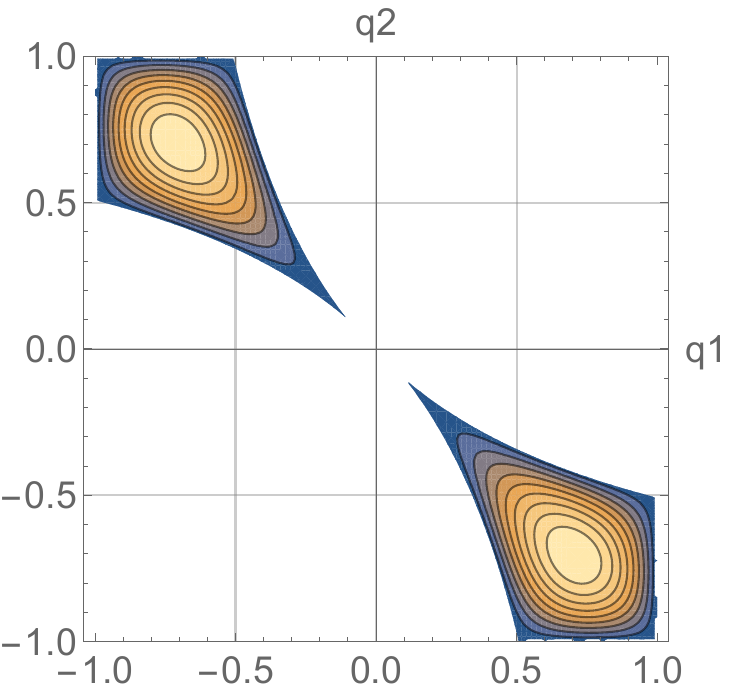}
 	\includegraphics*[width=0.07\columnwidth]{Stability-FODO-Barlegend.pdf}
	\caption{Contour plot of the stability of a FODO lattice. The white area is unstable. Shown are the stability in the $x$-plane (top), the $y$-plane (middle) and the stability of the complete system (bottom). As can be seen, the quads need to have opposite sign. We use the function $S_{a,b}^*(q_1,q_2,0)=4S_{a,b}(q_1,q_2,0)(1-S_{a,b}(q_1,q_2,0))$, which is also stable between zero and one, but is most stable when its value approaches one.}
	\label{fig:FODO-stability}
\end{figure}
\begin{figure}[!t]
	\centering
	 \includegraphics*[width=0.9\columnwidth]{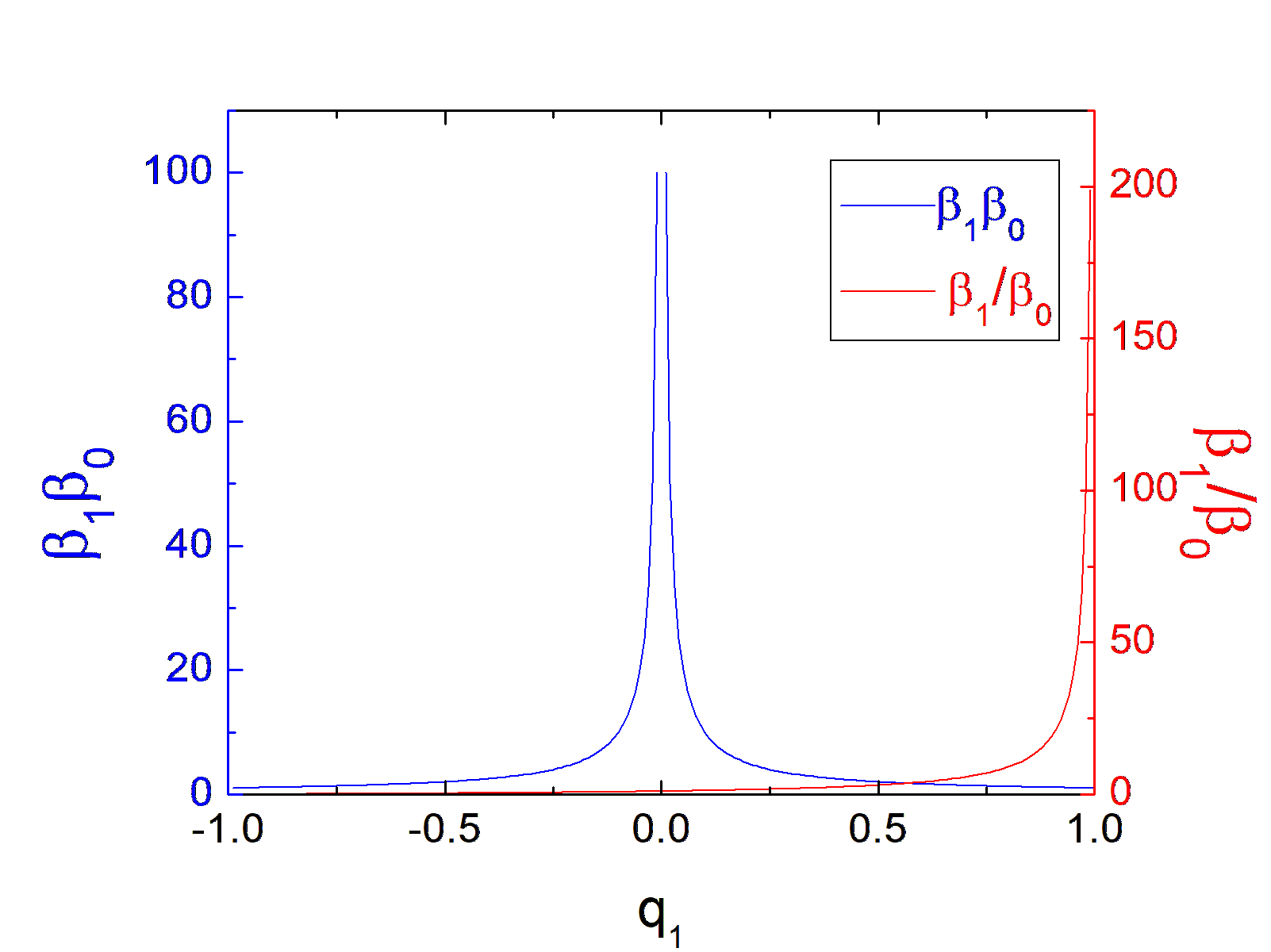}
	\caption{Average (blue curve and left scale) and variation (right scale with red curve) of the $\beta$-function of a FODO lattice, assuming opposite quads of equal strength. }
	\label{fig:FODO-beta}
\end{figure}
For a FODO-lattice ($\xi=0$), $A_{a,b}(\xi)=B_{a,b}(\xi)=1$, which means that the stability $S_{a,b}(\pm q_1,\pm q_2,\xi)$ in the two planes is rather straightforward $0\le (1\pm q_1)(1\pm q_2)\le 1$, where the positive and negative signs are used in the horizontal and vertical planes, respectively. This stability is shown in Fig.~\ref{fig:FODO-stability}. It is clear, that in order for both planes to be stable, $q_1$ and $q_2$ need to have opposite signs and the amplitude needs to be between $\pm 1$. The most common choice is to have both amplitudes equal, but if the focusing is strong enough $(|q_i|$ close to 1), also unequal focusing is possible. Instead of using the stability functions in Eq.~\eqref{Eq:stability2}, we use $4 S_{a,b}(q_1,q_2,0)(1-S_{a,b}(q_1,q_2,0))$ instead, which again needs to be between 0 and 1 in order to be stable, but is most stable for values approaching unity.

Using quads of equal, but opposite strength, results in the $\beta$-function as shown in Fig.~\ref{fig:FODO-beta}. For a small average $\beta$-function, the variation becomes large, whereas for a small variation, the average becomes large. As a compromise, $q_{1,2}=\pm 0.5$ is often chosen, which results in a variation of $\beta$ of a factor 3 and an average $\beta=2 l $. This is what is assumed as starting point in the remainder of this paper. Because we will assume planar, vertically focusing undulators, the stability properties remain unchanged in the horizontal plane with standard settings.

\subsection{Additional focusing with a thin planar undulator}

\begin{figure}[!t]
	\centering
	 \includegraphics*[width=\columnwidth]{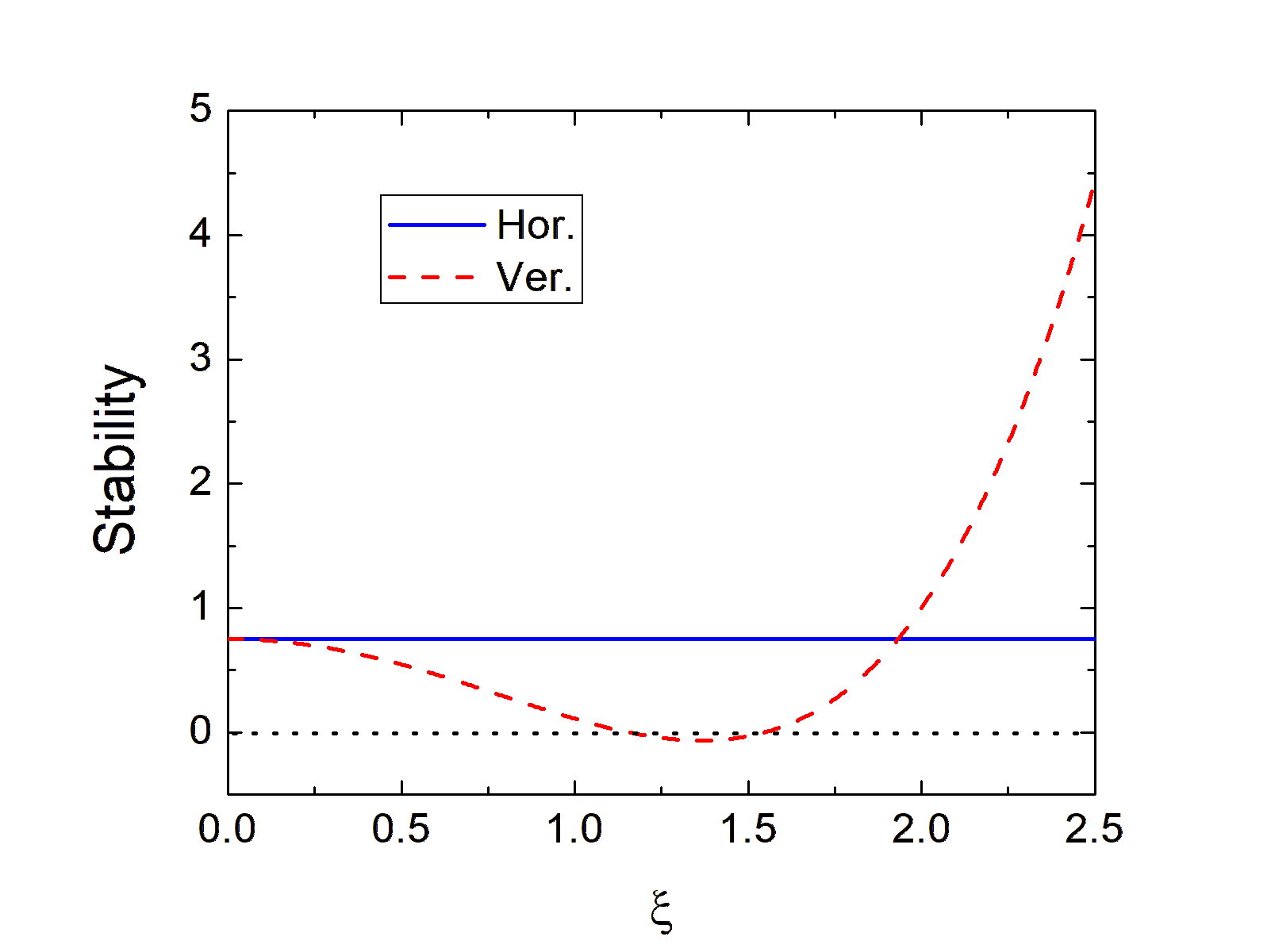}
	\caption{Stability of the system in the vertical plane with a zero-length undulator for $q_1=-q_2=0.5$. For this plot we have chosen $b=0$, which means that the undulator is centered between the two quads. There are only two small regions from 0 to $~1.2$ and $~1.5$ to 2 where the system is stable.}
	\label{fig:thin-undulator-stability}
\end{figure}

For a thin undulator, the same same approximation as for a thin quad is used, with $a=0$. In the vertical plane, $A_{0,b}(\xi)$ and $B_{0,b}(\xi)$ become
\begin{equation}
A_{0,b}(\xi)=1-\left[\frac{1}{4}-b^2\right]\xi^2\ \ {\rm and}\ B_{0,b}(\xi)=1-\left[\frac{1}{2}+b\right]\xi^2
	\label{Eq:thin-undulator-AB}
	\end{equation}
The total stability in the focusing plane of the undulator plane is therefore
\begin{widetext}
\begin{equation}
0\le \left(1-\frac{\xi^2}{2}+\left[1-\left(\frac{1}{4}-b^2\right)\xi^2\right]q_1\right)\left(1-\frac{\xi^2}{2}+\left[1-\left(\frac{1}{4}-b^2\right)\xi^2\right]q_2\right)\le 1
	\label{Eq:thin-undulator-stability}
	\end{equation}
\end{widetext}
The stability in Eq.~\eqref{Eq:thin-undulator-stability} is shown in Fig.~\ref{fig:thin-undulator-stability}. It is unstable almost everywhere, with the exception of two small regions. At $\xi=\pm 2/\sqrt{1-4b^2}$, where $A_{0,b}(\xi) =0$ and the quads have no influence on the system, the stability function is reduced to
\begin{equation}
S_{0,b}(q_1,q_2,\xi)=\left(\frac{1+4b^2}{1-4b^2}\right) \ge 1\,,
\end{equation}
which is always unstable, except for $b=0$, as stated before. When  $\xi=\sqrt{2}$ ($B_{0,0}(\xi)=0$), the stability function becomes
\begin{equation}
S_{0,b}(q_1,q_2,\xi)=\left(\frac{1}{2}+2b^2\right)^2q_1q_2 \,,
\end{equation}
the system can only be stable when $q_1$ and $q_2$ have the same sign and horizontally focusing. Because the coefficient is always between $1/4$ and $1$, this point can be made stable for $q_1$ and $q_2$ smaller than one, which guarantees stability in both planes. The region where this system is stable can be increased by adjusting the quad-settings.  Because it is not a realistic system, we do not attempt to improve its stability.

\subsection{Additional focusing with a planar undulator filling the complete space}

\begin{figure}[!t]
	\centering
	 \includegraphics*[width=\columnwidth]{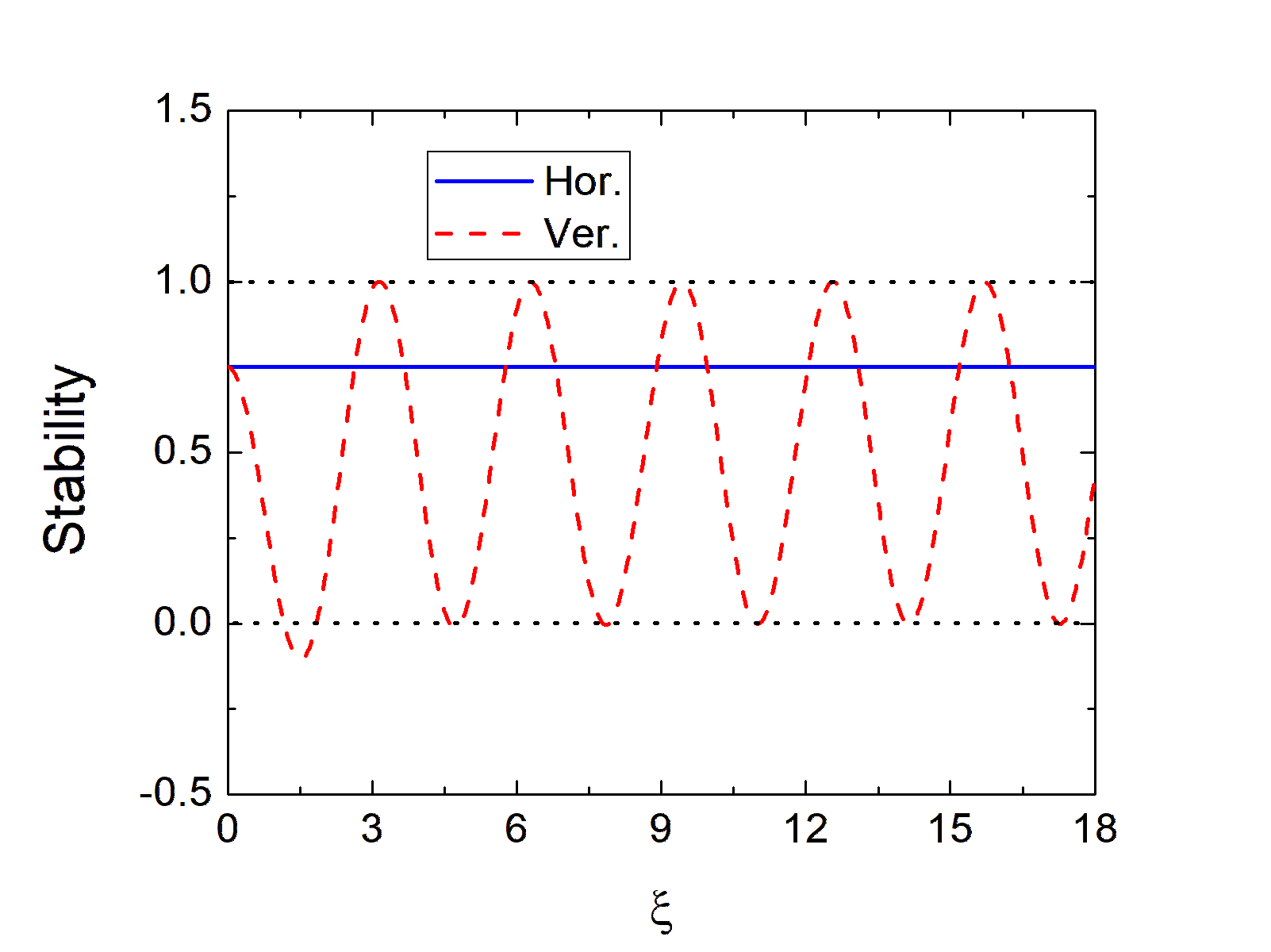}
	 \includegraphics*[width=\columnwidth]{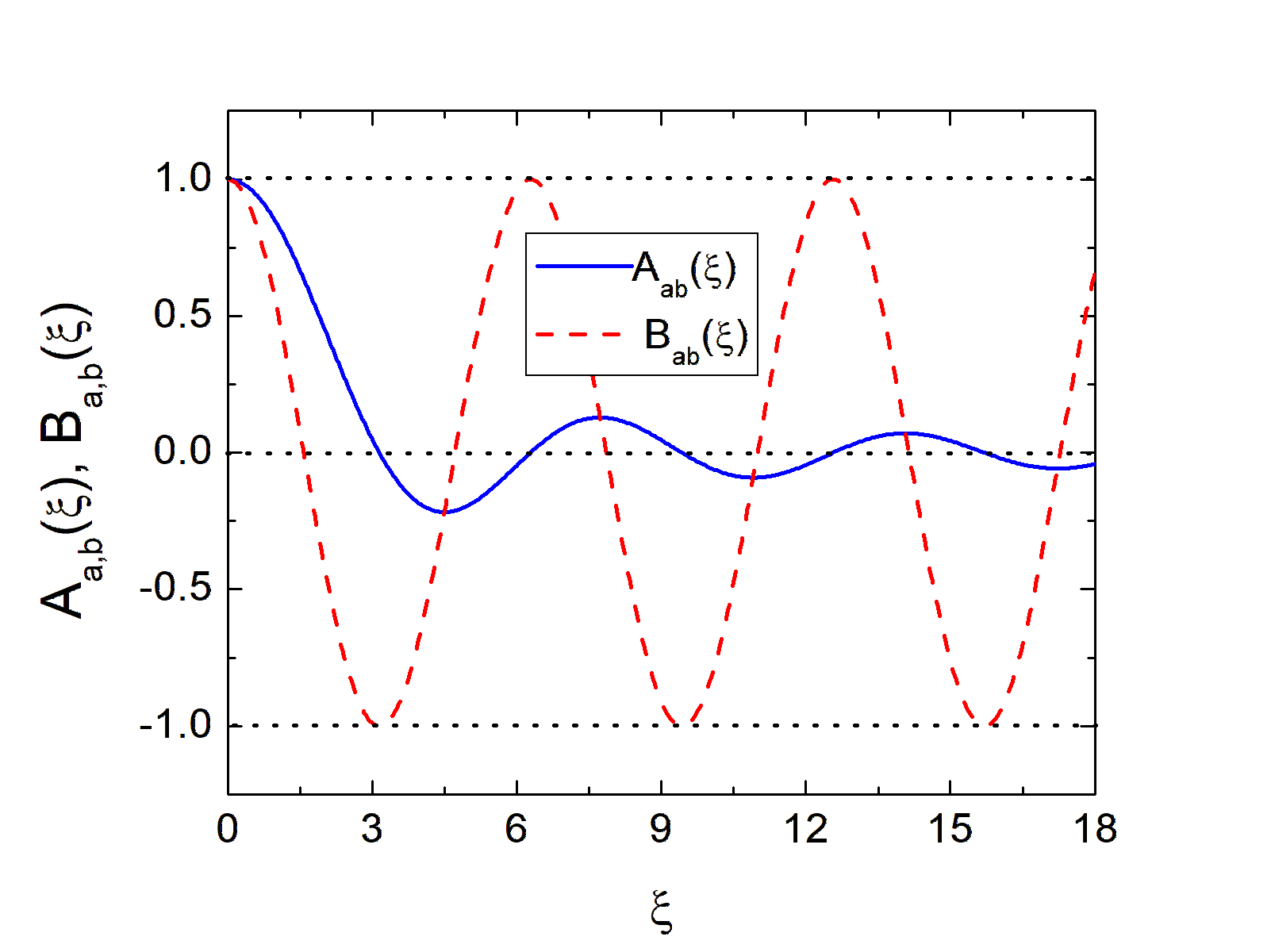}
	 	 	\caption{Stability function with $q_1=-q_2=0.5$ (top) in the horizontal (red) and vertical (blue dashed) plane and the behavior of functions $A_{1,0}(\xi)$ in red and $B_{1,0}(\xi)$ in blue as a function of $\xi$ (bottom). The maximum of the oscillating stability function is one, and therefore stable, the minima are all below zero, and therefore unstable, but approach zero as $\xi$ increases. As can be seen, the function $A_{1,0}(\xi)$ has zeros at multiples of $\pi$ and is an oscillating function with reducing amplitude whereas the function $B_{1,0}(\xi)$ has zeros exactly between them and has constant amplitude.}
	\label{fig:full-undulator-stability}
\end{figure}
Closer to reality is a planar undulator that fills the complete space. This means that $a=1$ and therefore automatically $b=0$.
\begin{equation}
A_{1,0}(\xi)=\frac{\sin\xi}{\xi}\ \ {\rm and}\ B_{1,0}(\xi)=\cos\xi
\end{equation}
\begin{figure}[!t]
	\centering
	 \includegraphics*[width=\columnwidth]{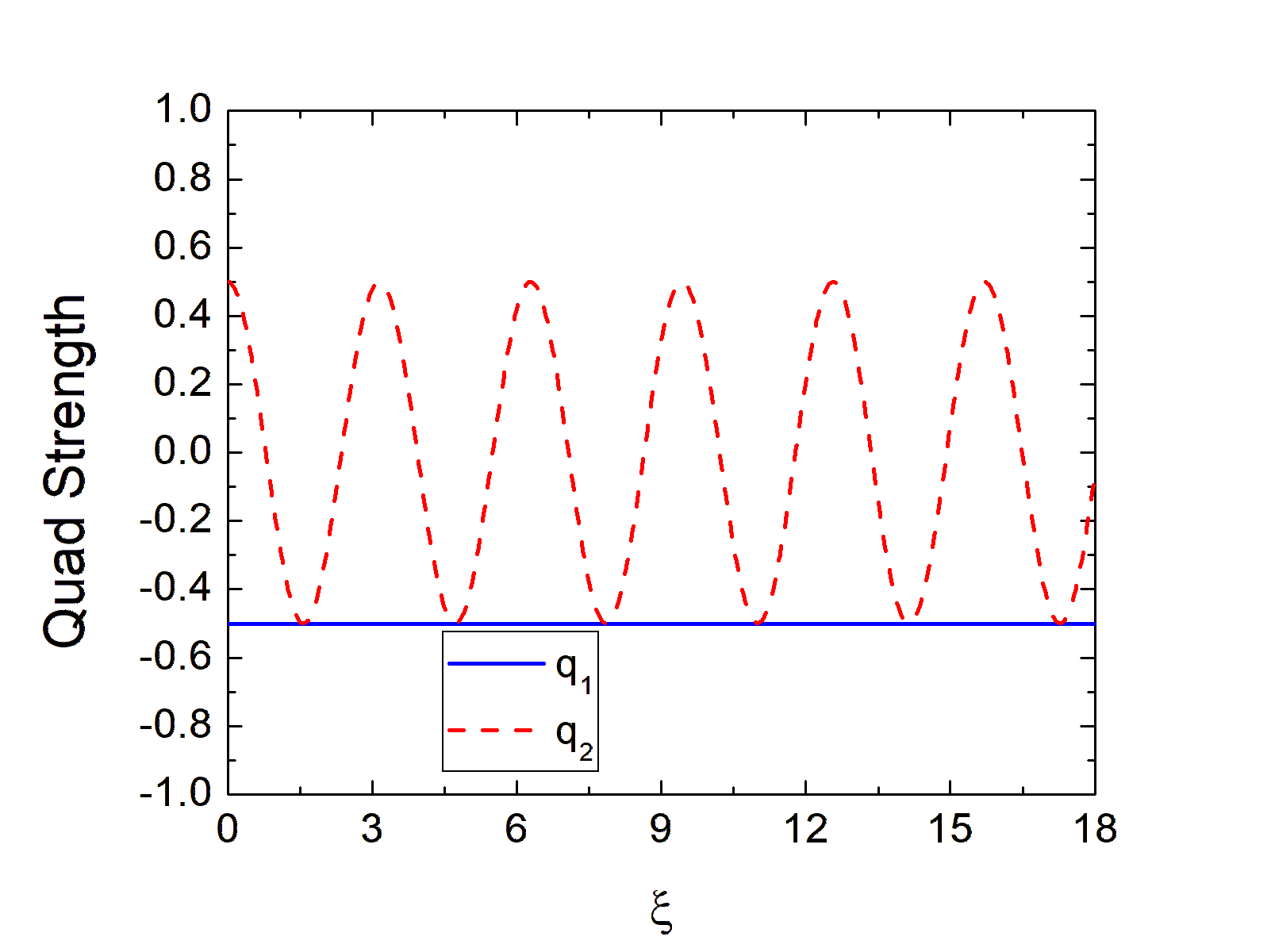}
	 	 \includegraphics*[width=\columnwidth]{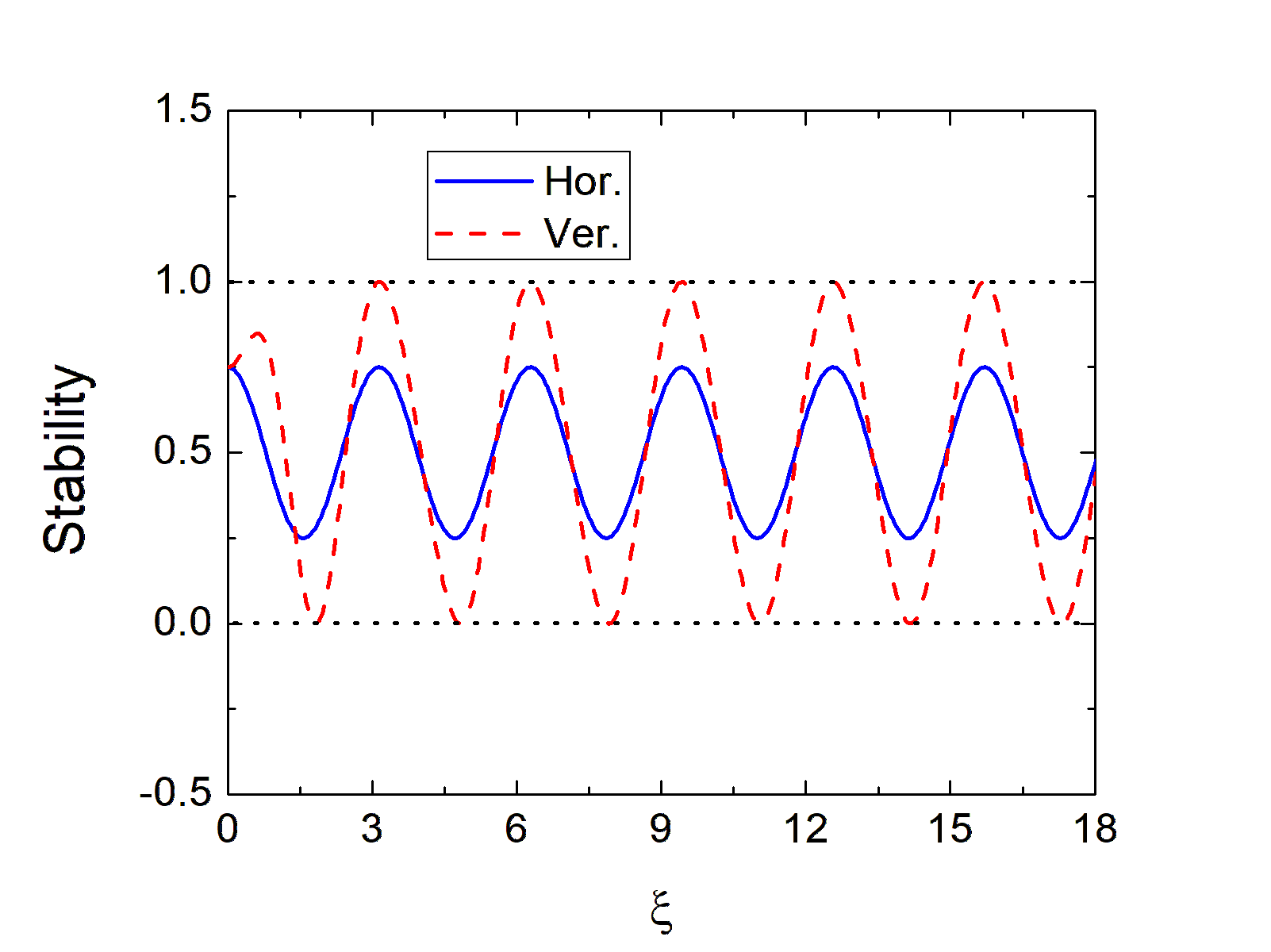}
	\includegraphics*[width=\columnwidth]{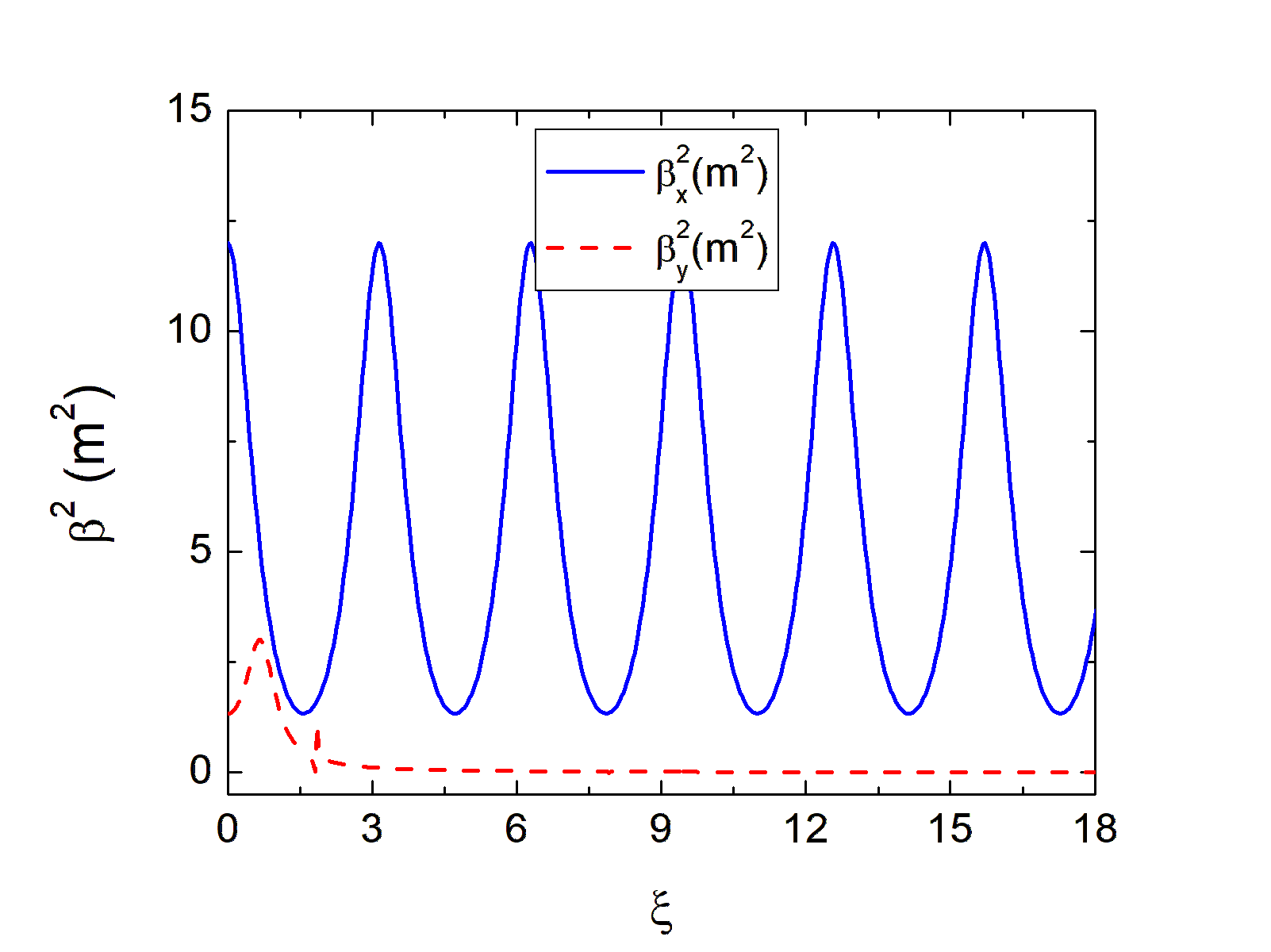}
	 	\caption{Quad settings to increase stability (top with $q_1$ in red and $q_2$ in blue), Stability function with adapted quad settings (middle, with red for the horizontal and blue for the vertical plane) and Value of $\beta$-function in both planes (bottom with again red for the horizontal and blue for the vertical plane.). }
	\label{fig:full-undulator-stability-adjusted}
\end{figure}
The stability $S_{1,0}(q_1,q_2,\xi)$ becomes

\begin{equation}
0\le  \left( \cos \xi +{\sin}{\rm c}\,\xi q_1\right )\left ( \cos \xi +{\sin}{\rm c}\,\xi q_2 \right) \le 1
\end{equation}

The stability in both planes and the functions $A_{1,0}(\xi)$ and $B_{1,0}(\xi)$ are shown in Fig.~\ref{fig:full-undulator-stability}. Compared to the thin-undulator approximation, the functions $A_{1,0}(\xi)$ and $B_{1,0}(\xi)$ become oscillating functions of $\xi$ and the number of special points becomes infinite.
$A_{1,0}(\xi)=0$ when $\xi=n\pi$, where $n$ is a non-negative integer. In this case the stability becomes $S_{1,0}(q_1,q_2,n\pi)=1$, which is always stable. The other set of special points is when $B_{1,0}(\xi)=0$, which is the case when $\xi=(n+1/2)\pi$. In this case,
\begin{equation}
S_{1,0}(q_1,q_2,\xi)=\frac{q_1q_2}{\left((n+1/2)\pi\right)^2}\,,
\end{equation}
\begin{figure}[!t]
	\centering
	 \includegraphics*[width=\columnwidth]{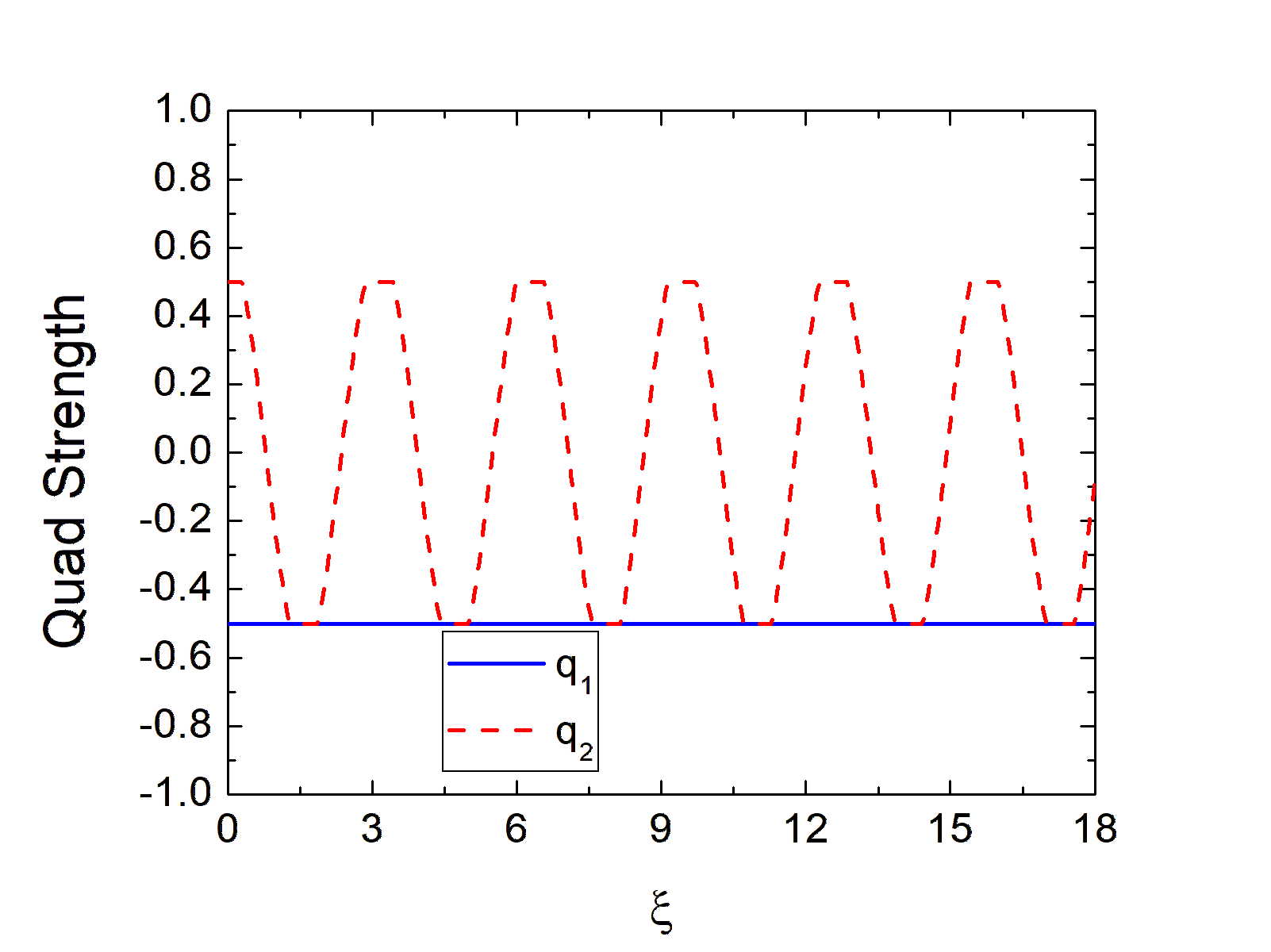}
	 \includegraphics*[width=\columnwidth]{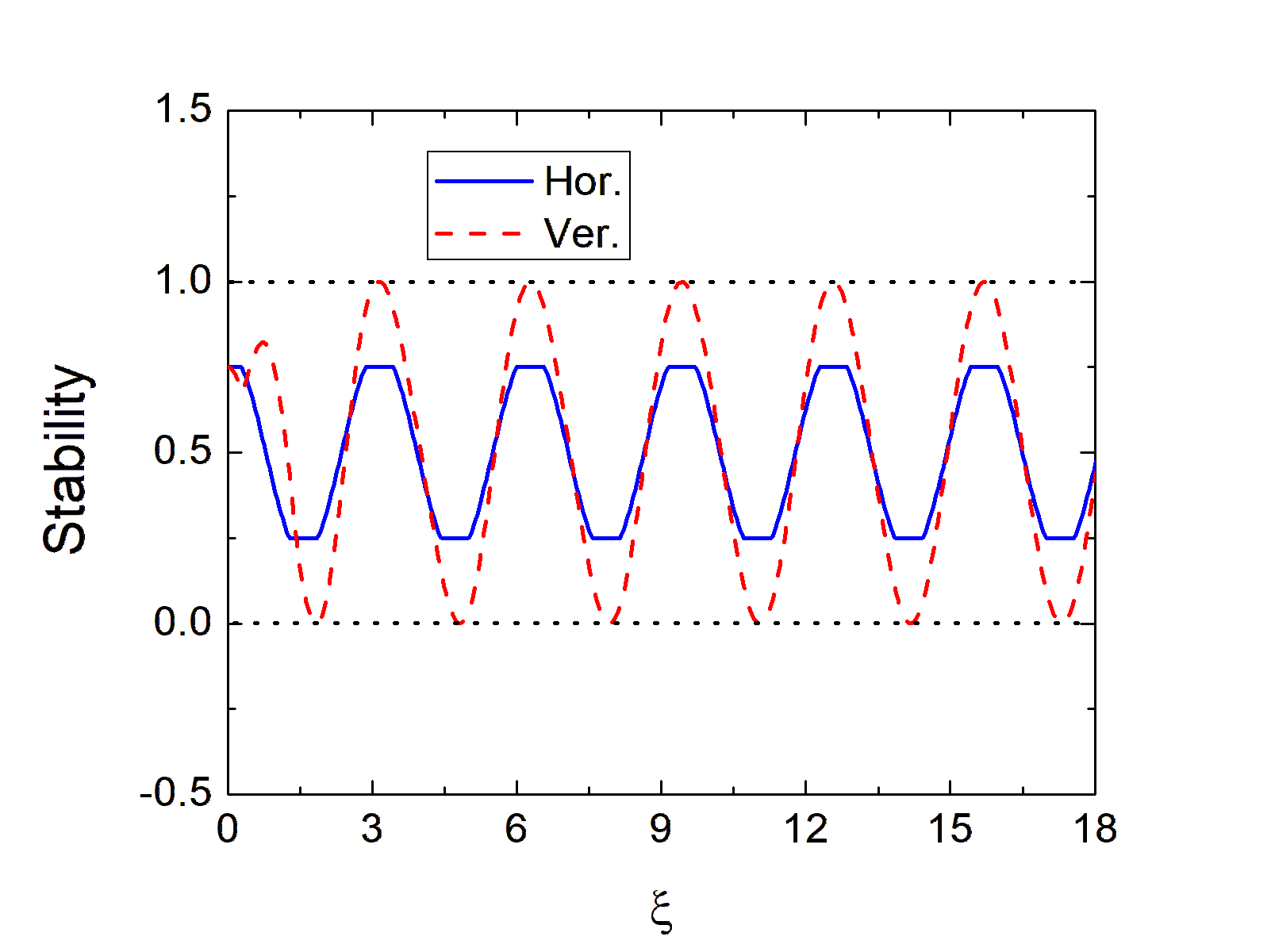}
	\includegraphics*[width=\columnwidth]{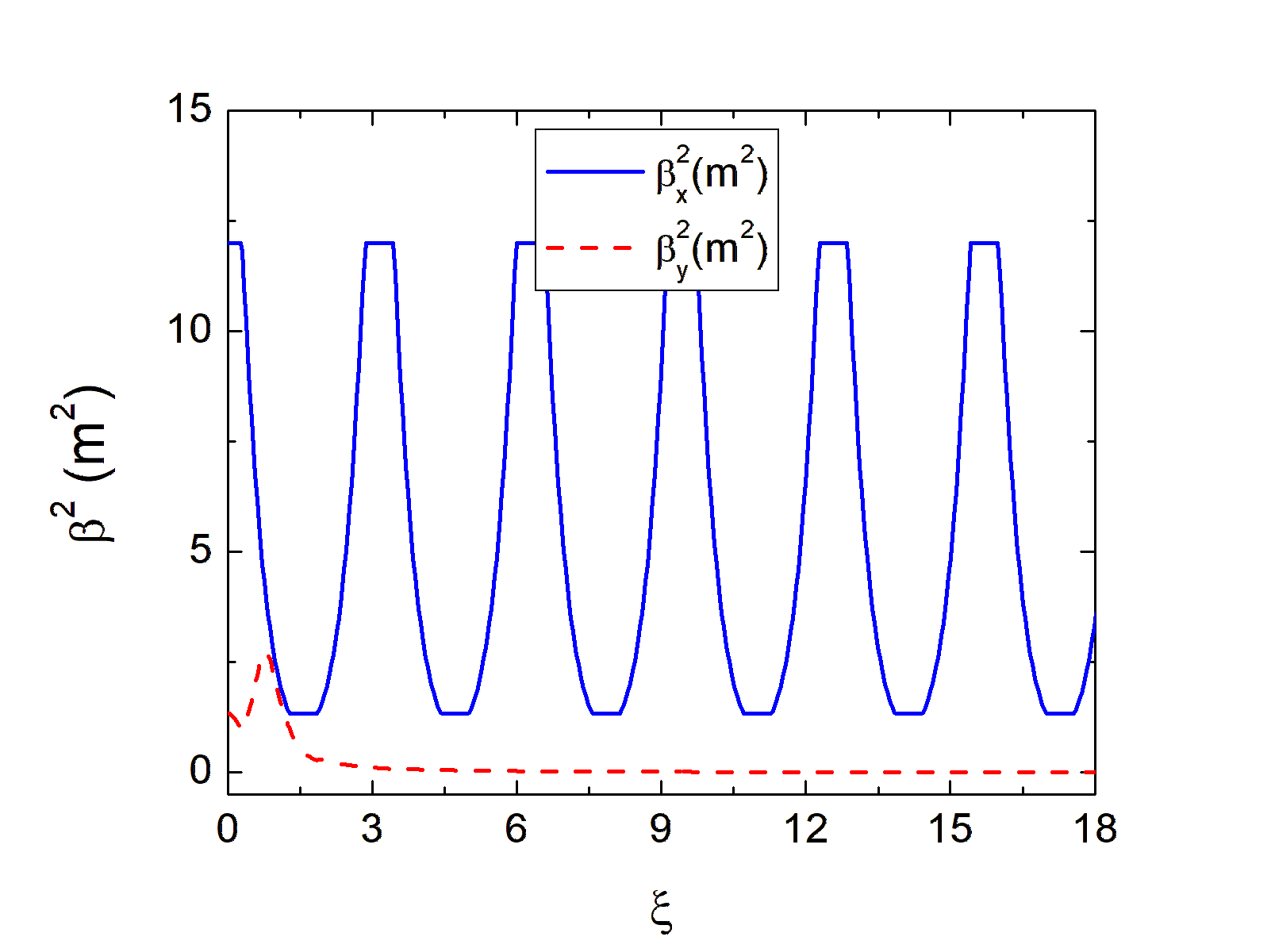}
	 	\caption{Quad settings to increase stability (top with $q_1$ in red and $q_2(\xi)$ in blue), Stability function with adapted quad settings (middle, with red for the horizontal and blue for the vertical plane) and value of $\beta$-function in both planes (bottom with again red for the horizontal and blue for the vertical plane.). }
	\label{fig:full-undulator-stability-adjusted2}
\end{figure}
In order to make $S_{1,0}(q_1,q_2,\xi)$ stable everywhere, one can choose  $q_1=-1/2$ as before, but for $2q_2=\cos(2\xi)$. This means that the quad-focusing changes from a FODO structure at $\xi=n\pi$ to a FOFO structure with equal amplitude at $\xi=(n+1/2)\pi$. The stability is shown in Fig.~\ref{fig:full-undulator-stability-adjusted}. The top shows the variation in quadrupole strength, the middle the stability in both planes and the bottom the value of $\beta^2$ at the beginning of the cell in both planes. It is not  entirely stable, as can be seen at $\beta^2$ around $\xi=\pi/2$. The reason for this is that the explicit dependency of $q_2(\xi)$ on $\xi$ changes the stability function, which moves the value of $\xi$ for which the stability reaches its minimum. A truncated oscillating function for $q_2(\xi)$, as shown in Fig.~\ref{fig:full-undulator-stability-adjusted2} avoids this problem.

In principle, any function can be used to make the system stable. The above function for $q_1$ and $q_2$ ensures smooth transitions, which makes for example wavelengths scans performed by changing undulator gap or beam energy much easier to perform. A sudden change in quad-setting will cause strong change in divergence of the photon beam and in general also cause orbit changes. A smooth transition will allow feedbacks to act and partially compensate for unwanted changes.

Even though the behavior of this system is more complicated than the ``thin undulator'' version before, it is rather straightforward to make the system stable. However, in practice, space is needed for diagnostics and phase-shifters. Therefore, it is not realistic to assume that the complete space between the (thin) quads is filled with undulators. The next section will study a system including realistic drift paces around the quadrupoles.

\section{Model used for a realistic planar undulator}

\begin{figure}[!t]
	\centering
	 \includegraphics*[width=\columnwidth]{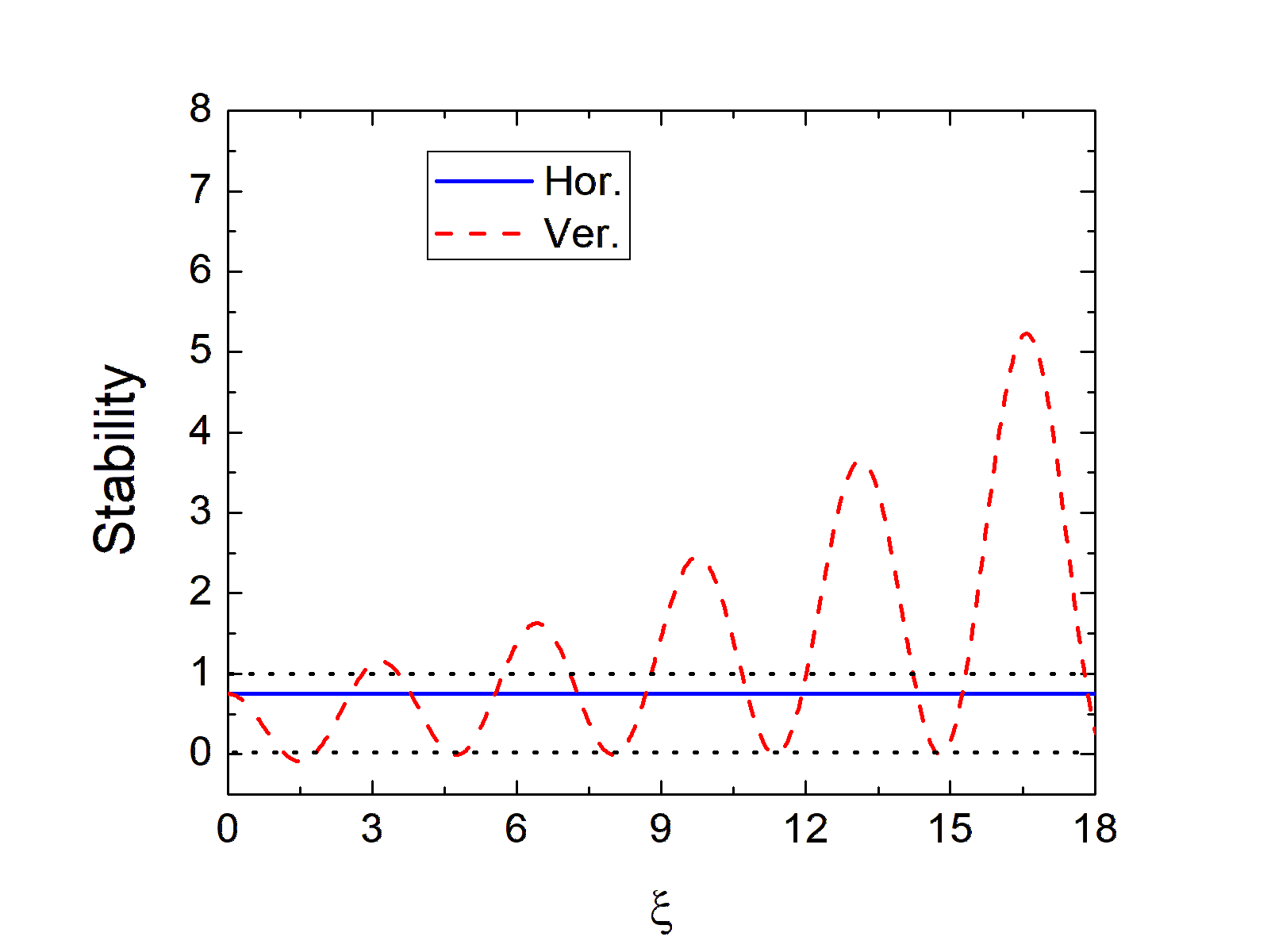}
	 \includegraphics*[width=\columnwidth]{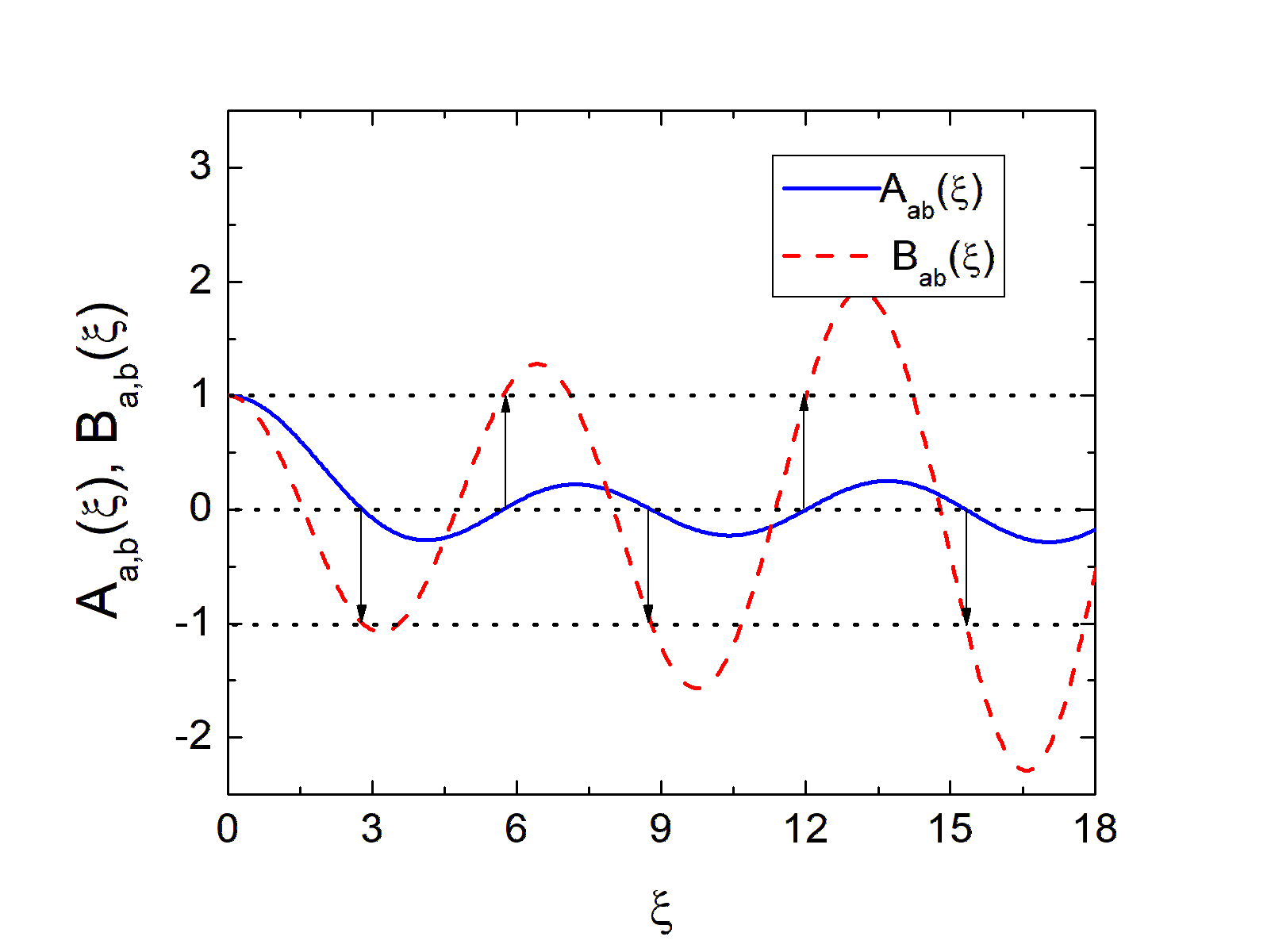}
	 	\caption{Stability function with $q_1=-q_2=0.5$ (top) in the horizontal (red) and vertical (blue dashed) plane and the behavior of functions $A_{a,b}(\xi)$ in red and $B_{a,b}(\xi)$ in blue as a function of $\xi$ (bottom). The maximum of the oscillating stability function grows for larger values of $\xi$ to values larger than 1, and therefore unstable, the minima are all below zero, and therefore unstable, but approach zero as $\xi$ increases. As can be seen, the function $A_{a,b}$ has zeros close multiples of $n$ and is an oscillating function with reducing amplitude whereas the function $B_{a,b}$ has zeros approximately between them and has increasing amplitude. For this figure, $a=0.8$ and $b=0$ has been chosen.}
	\label{fig:real-undulator-stability}
\end{figure}

It is easier to stabilize for an undulator filling the complete space than for a thin undulator with zero length. On the other hand, the functions $A_{1,0}(\xi)$ and $B_{1,0}(\xi)$ have become considerably more complicated and have many zero's. For the real system, with an undulator between two quads with drift spaces around them, the stability and behavior of the functions  $A_{a,b}(\xi)$ and $B_{a,b}(\xi)$ is shown in Fig.~\ref{fig:real-undulator-stability}, assuming $a=0.8$ and $b=0$. The corresponding equations are complicated and are discussed in the appendix. Except that the system is only stable when $b=0$ for $A_{a,b}(\xi)=0$, more general statements are difficult to give.

\begin{figure}[!t]
	\centering
	 \includegraphics*[width=0.95\columnwidth]{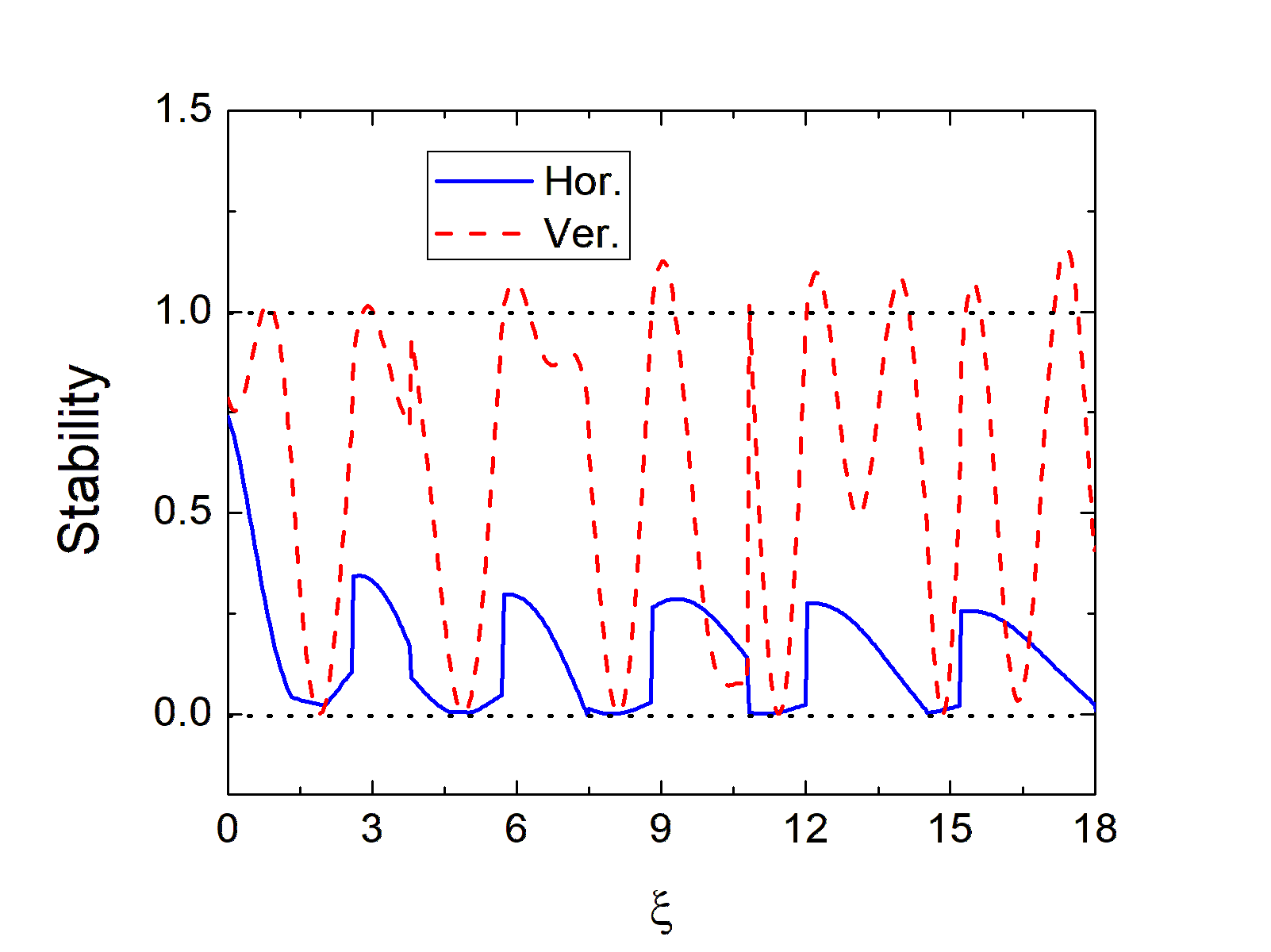}
	 \includegraphics*[width=0.95\columnwidth]{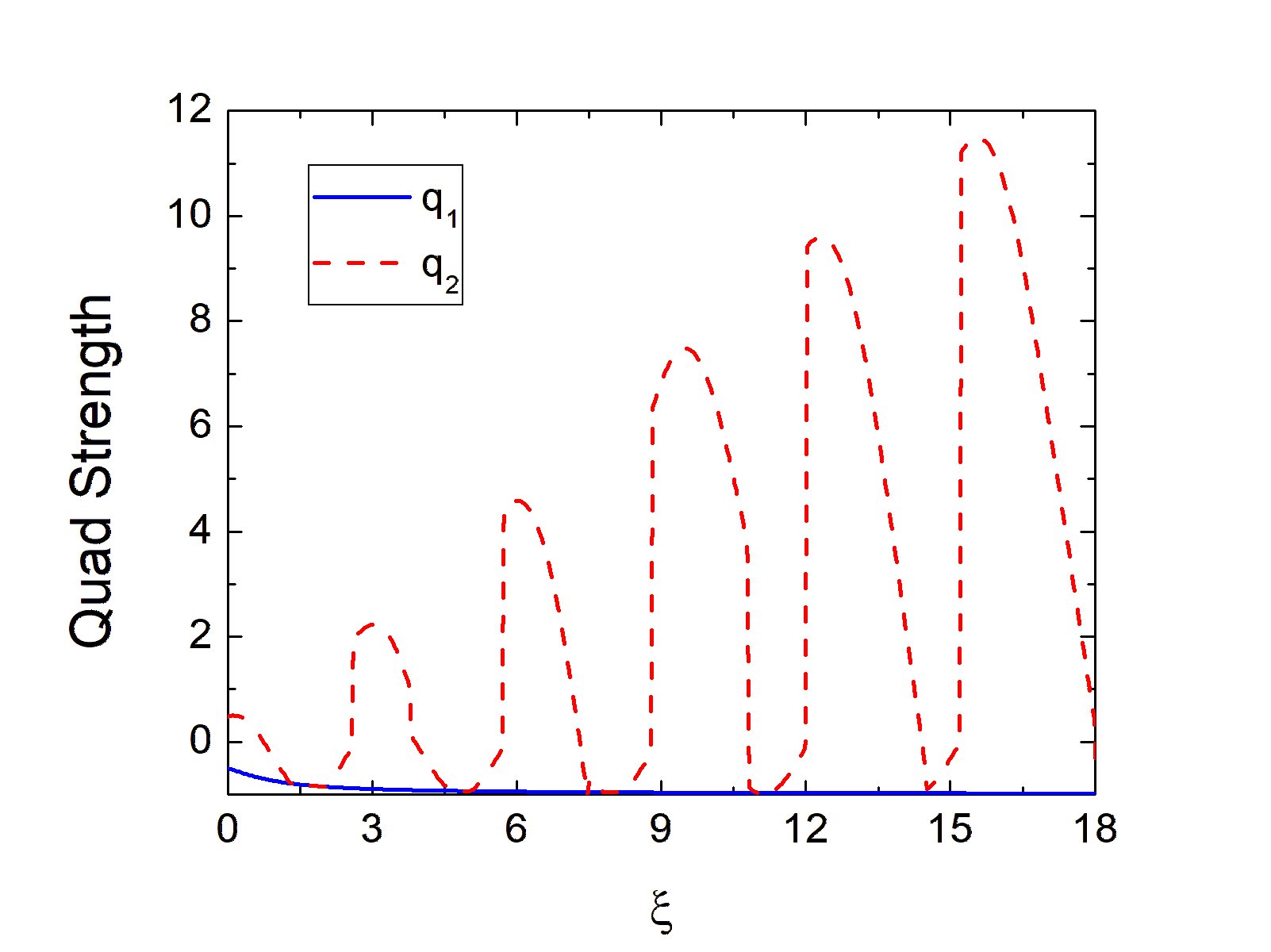}
	 	\caption{Stability function using standard settings (top) and step functions for the quad focusing (bottom).  For this figure, $a=0.8$ and $b=0$ has been chosen. }
	\label{fig:real-undulator-stability-corrected}
\end{figure}

In general, because of the complicated stability function, it is hard to find a general function for $q_1$ and $q_2$ that is stable everywhere. In order to compensate for the stronger focusing of the undulator with increasing $\xi$, the defocusing by the quads needs to increase as well. However, this causes problems in the other plane, which can only be avoided by choosing the other value for the quad close to $-1$, such that $0\le (1+q_1)(1+q_2)\le1$ remains true. This means that $q_1$ can be a function that starts at $-1/2$ and approaches $-1$ for large $\xi$, for example $q_1=-(1+\arctan(\sqrt{a}\xi)/2$. This makes the horizontal plane stable, even for large $\xi$, while $q_2$ grows to large values. In the undulator (vertical) plane, there is no simple function, which means we try a set of step functions instead, which is shown in Fig.~\ref{fig:real-undulator-stability-corrected}. As can be seen, the $q_2(\xi)$-function is locally fitted with a parabolic profile, as shown in bottom of Fig.~\ref{fig:real-undulator-stability-corrected}, while $q_1(\xi)=-(1+\arctan{\xi}/\pi)/2$. The stability of the system, shown at the top of Fig.~\ref{fig:real-undulator-stability-corrected} is not entirely stable, but one can imagine that with further optimization, this could achieved. Further study is ongoing, to find a general theoretical proof of stability. In practice, a normal undulator system will hardly reach these large values of $\xi$, which is why in the next section, the parameters of SXFEL will be studied in detail.

\section{Stable focusing at SXFEL}

\begin{figure*}[!t]
	\centering
	 \includegraphics*[width=\textwidth]{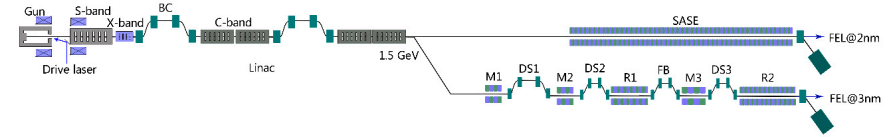}

	\caption{Layout of the SXFEL facility as test-facility (top) and user facility (bottom). parameters are given in Table~\ref{table1}.}
	\label{fig:Layout-SXFEL}
\end{figure*}

As mentioned in the introduction, the issue of undulator focusing has been important for several FELs. In this case, we focus our attention to the SXFEL \cite{Zhao2017}. The layout is shown in Fig.~\ref{fig:Layout-SXFEL}. Parameters of the SBP undulator, which is the SASE-FEL, are given in Table~\ref{table:SBP}. For the SUD-FEL, which is the seeded FEL, the focusing of the undulator does not play a dominant role for the nominal energy range of the facility.
\begin{figure}[!t]
	\centering
	 \includegraphics*[width=0.8\columnwidth]{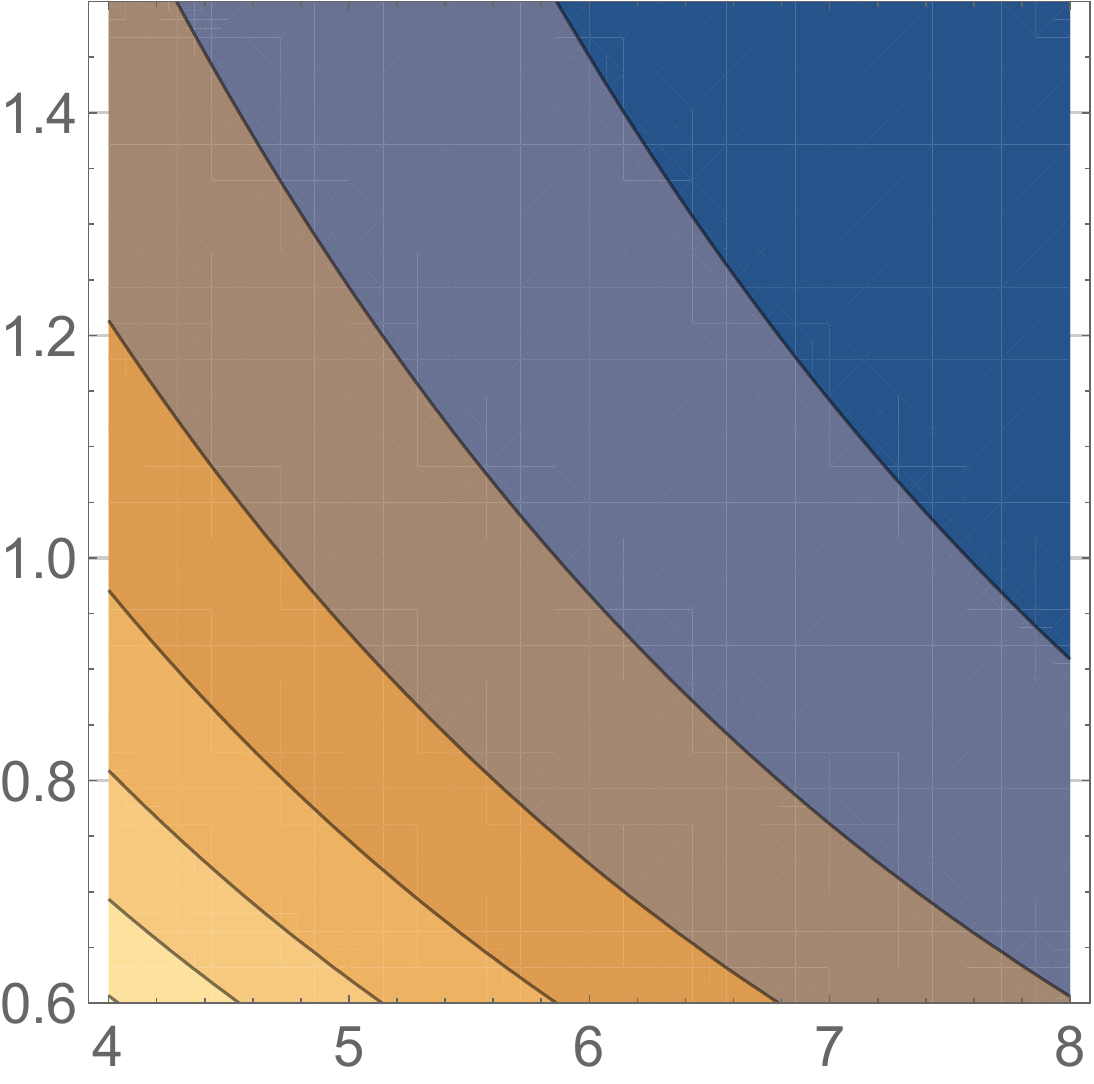}
	 	 \includegraphics*[width=0.085\columnwidth]{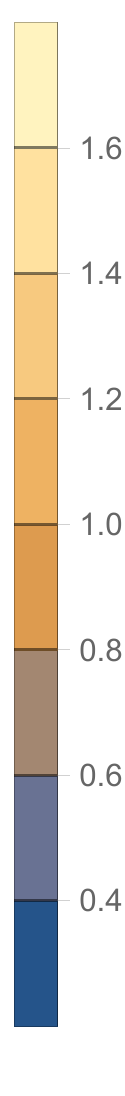}
	 	 	 \includegraphics*[width=0.8\columnwidth]{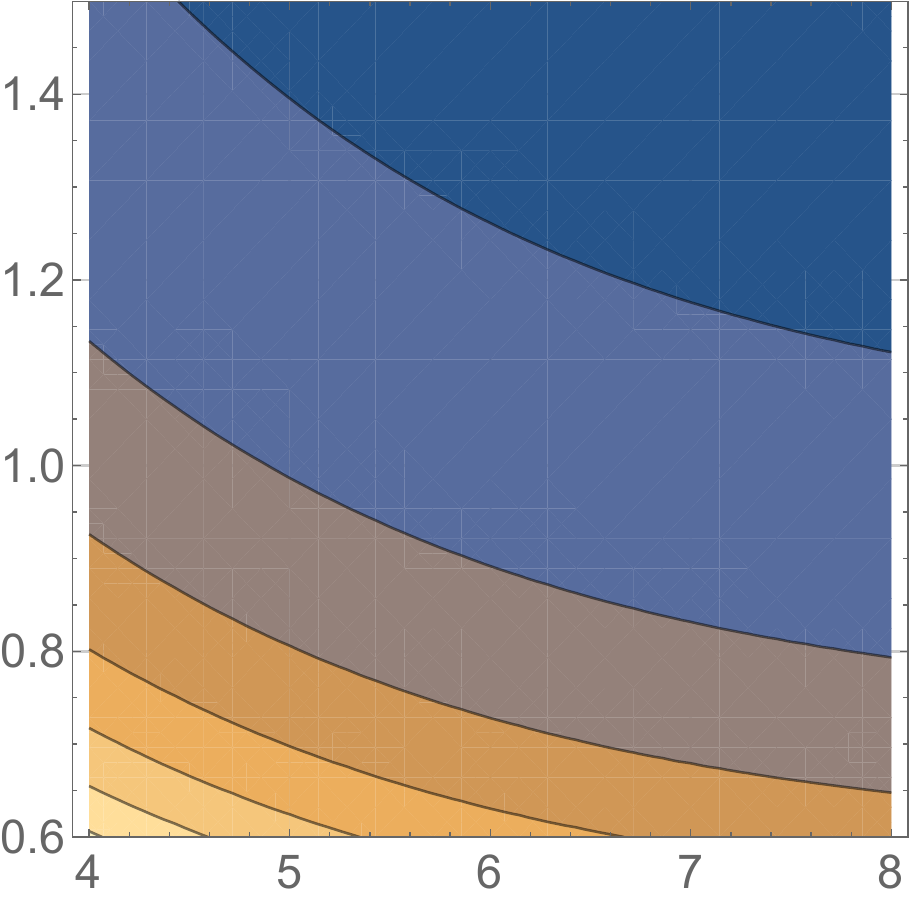}
	 \includegraphics*[width=0.085\columnwidth]{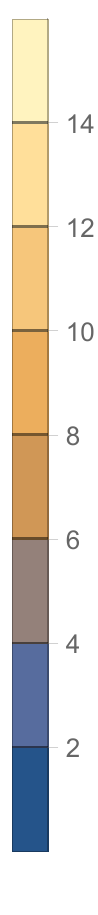}
	 	\caption{Limits on parameters for the SXFEL. The value of $\xi$ (top) and the maximum wavelength (in mm, bottom) as function of beam energy (GeV) and undulator gap (mm).}
	\label{fig:SXFEL-limits}
\end{figure}

 For the range of gaps  of the in-vacuum undulators, between 4 and 8~mm, and the possible energy range between 0.6 and 1.5~GeV, values of $\xi$ and the wavelength in nanometer are shown in Fig.~\ref{fig:SXFEL-limits}.  As can be seen in the top figure, even at the lowest energy and the minimum gap, the value of $\xi$ does not exceed 1.62.
\begin{table}[!t]
	\begin{center}
		\caption{parameters for the undulators of the SASE undulator at SXFEL-UF.}
		\label{table1}
		\begin{tabular}{l l l l}
			 & \textbf{Unit}  & \textbf{SBP}& \textbf{}\\
			\hline
			{\bf Electron beam} & &  & \\
			Energy range& GeV& 0.6 - 1.5 & \\
			Emittance & \si{\um}  & 1.5 &\\
			 & &  & \\
			 \hline
			{\bf Undulator} & &  & \\
			gap & mm & 4 - 8 & \\
			Period & mm & 16 & \\
			\# segments & & 10 & \\
			Length segment & m& 4 & \\
			Intersection length & m & 1.1  & \\
			Average beta function &m & 10 & \\
			 & &  & \\
			 & &  & \\

	         \hline
	\label{table:SBP}
			\end{tabular}
	\end{center}
\end{table}

\begin{figure}[!t]
	\centering
	 \includegraphics*[width=\columnwidth]{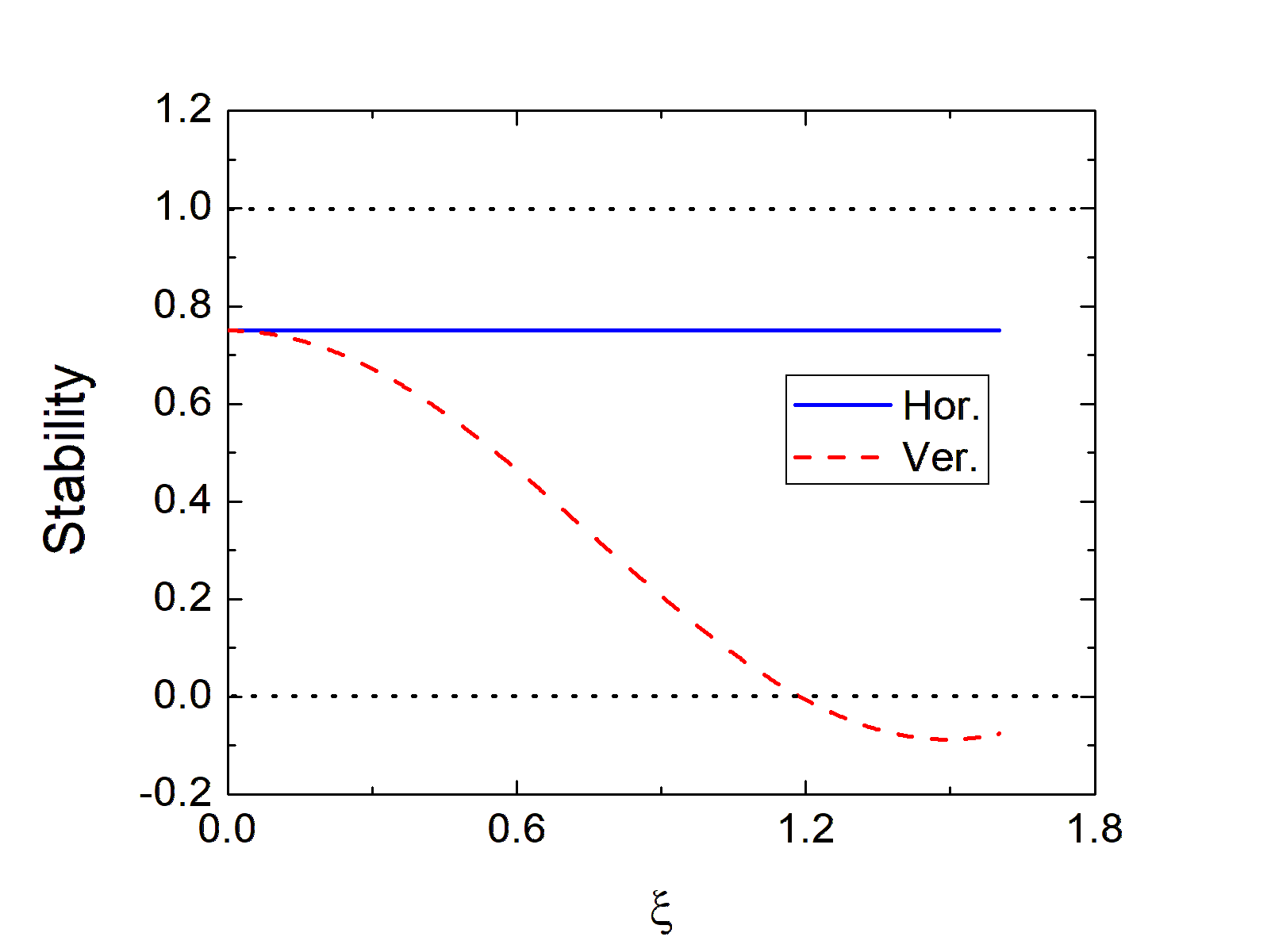}

	 	\caption{Stability of the SXFEL SBP undulator. As can be seen, it only covers the first instability and actually becomes stable again at the lowest energy at a fully closed undulator gap.}
	\label{fig:SXFEL-stability}
\end{figure}

\begin{figure}[!t]
	\centering
	 \includegraphics*[width=\columnwidth]{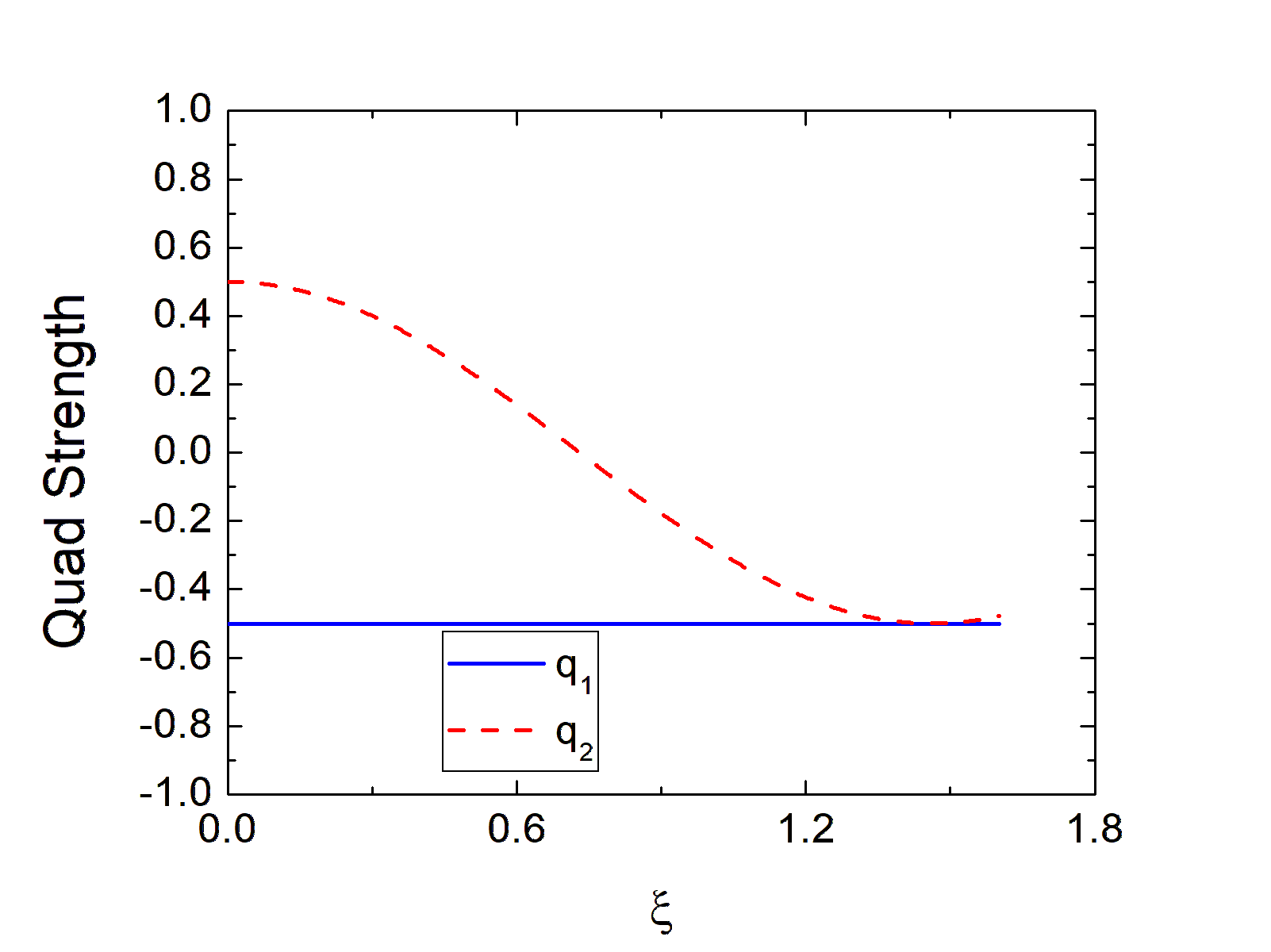}
	 \includegraphics*[width=\columnwidth]{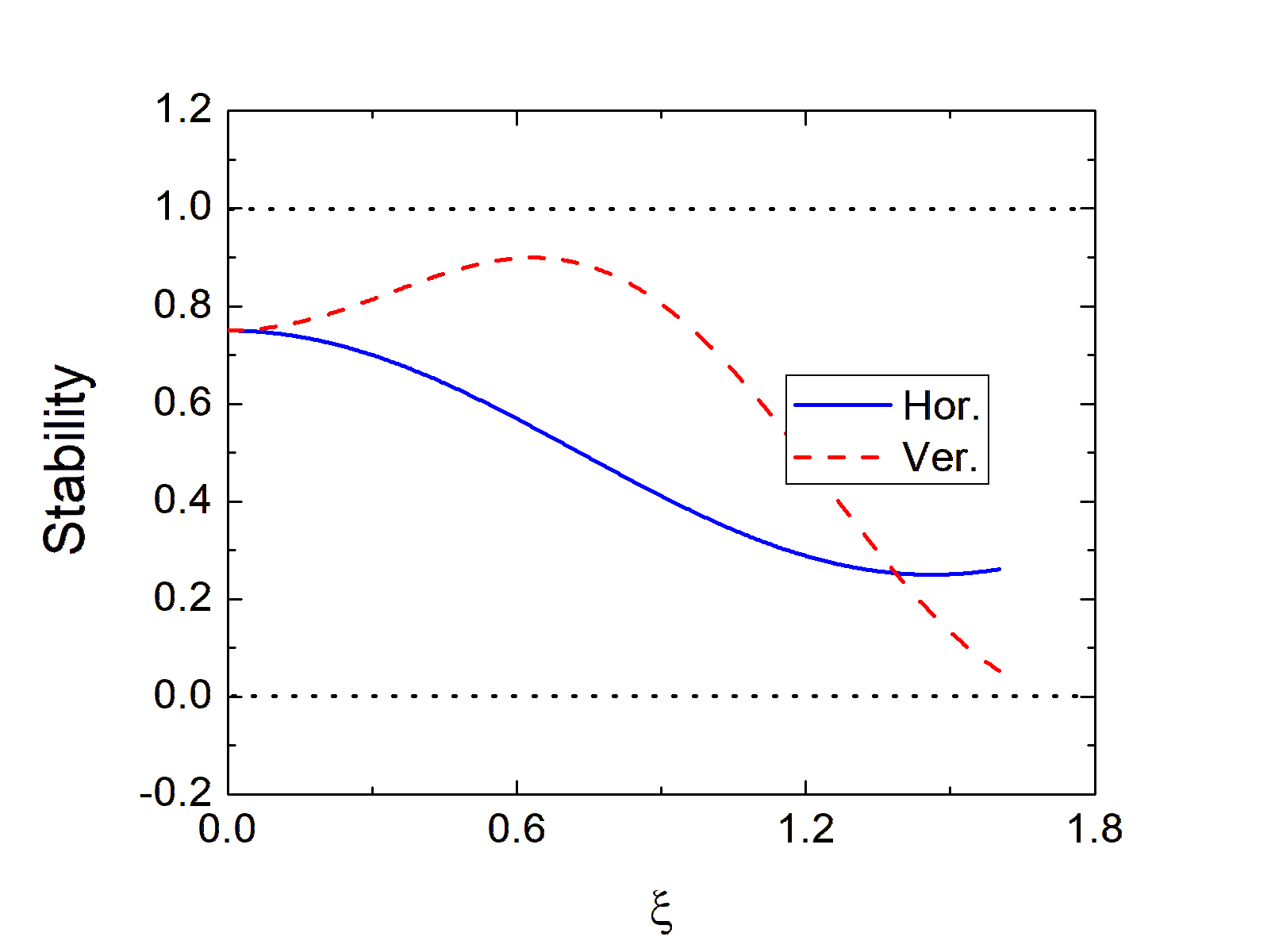}
	 	\caption{Modified, $\xi$-dependent quad settings (top) and stability for the SXFEL SBP undulator}
	\label{fig:SXFEL-stability-corrected}
\end{figure}

The stability function is shown in Fig.~\ref{fig:SXFEL-stability}. As can be seen, only the first instability region is covered. As for the general case, we again choose a $\xi$-dependent quad setting that was used for the undulator completely filling the space between the quads to modify the stability of the undulator focusing, as shown in Fig.~\ref{fig:SXFEL-stability-corrected}. This gives good results, but as was already mentioned, in principle, a completely different function can be chosen instead. Because there is no risk that any of the other resonances can be reached, a much simpler function will make the focusing stable everywhere, but would become unstable at larger values of $\xi$.

\begin{figure}[!t]
	\centering
	 \includegraphics*[width=\columnwidth]{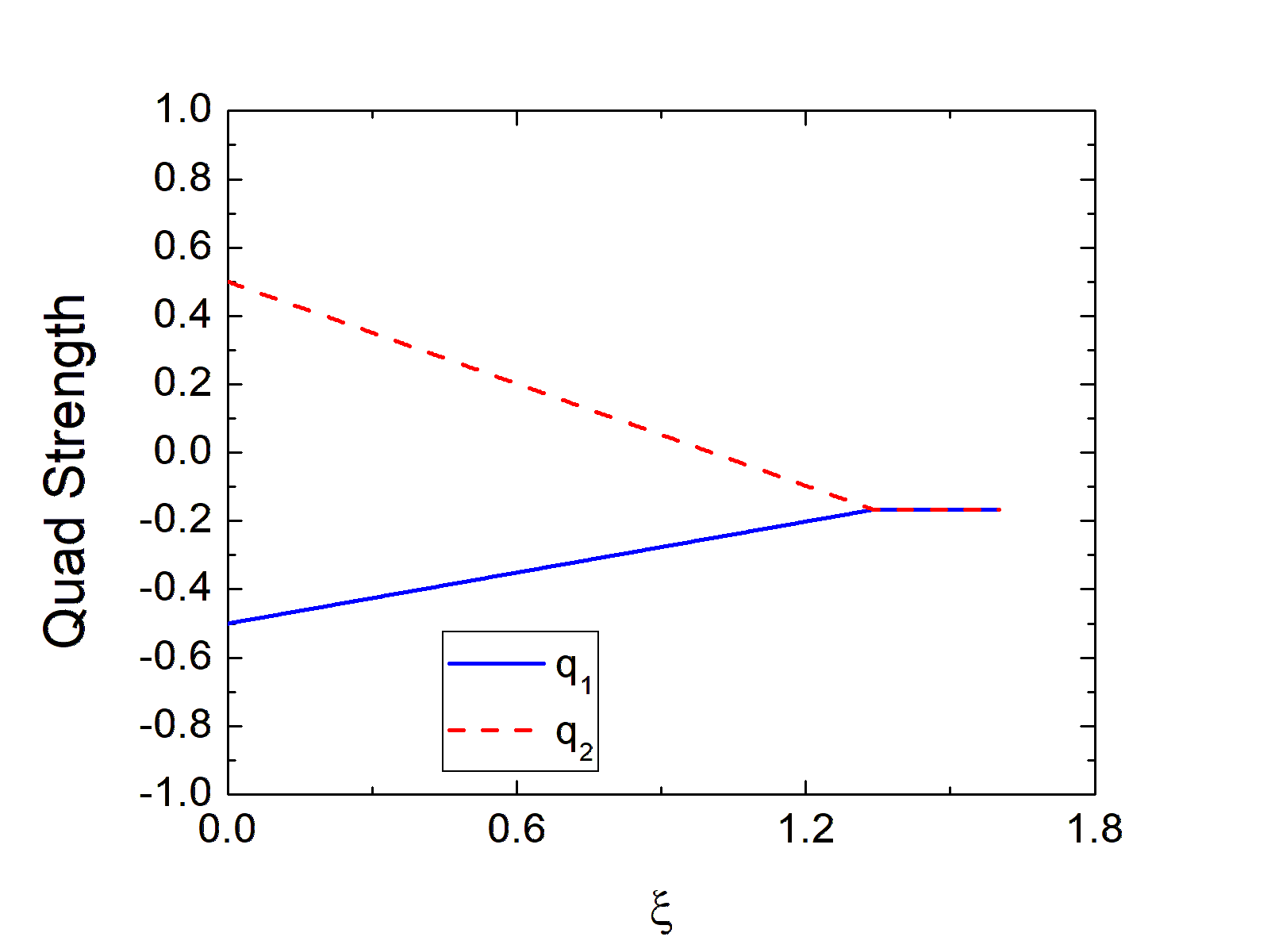}
	 \includegraphics*[width=\columnwidth]{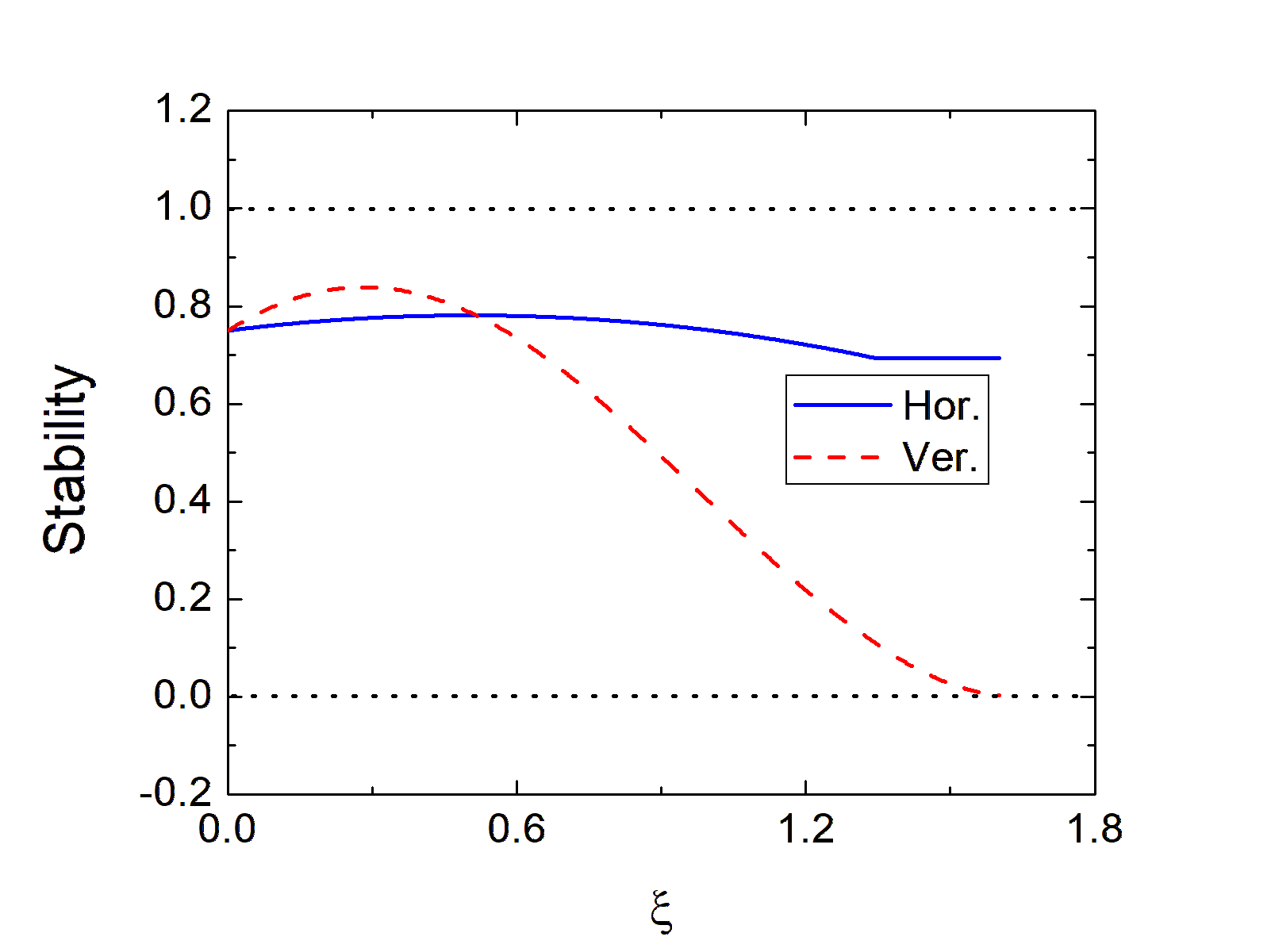}
	 	\caption{Modified quad settings with a linear dependency on $\xi$ (top) and stability for the SXFEL SBP undulator using these quad settings.}
	\label{fig:SXFEL-stability-corrected2}
\end{figure}

For FLASH2, a quadratic function was chosen for the $\xi$-dependence of the quads, where the coefficients were different for $q_1$ and $q_2$ to change from FODO to FOFO. Here, we assume a linear dependent $\xi$-function for both quads, where the strength of the quads needs to be exactly equal at the location where the stability function is zero. Because this minimum is moving with changes in $q_1$ and $q_2$, this minimum needs to be determined iteratively, which is rather straightforward. Also in this case, it is more stable to have a region where both quads have equal strength, which means a slight modification of the quad functions. The variation in quad strength and the corresponding stability function are shown in Fig.~\ref{fig:SXFEL-stability-corrected2}.

\section{Conclusion and discussion}

The automatic optics has been implemented and tested in the SASE beamline of the FLASH2. It has been shown experimentally that the automatic optics works and ensures that beam loss can be avoided for the entire wavelength range of this FEL. However,  number of conclusions can be drawn based on the findings in this paper
\begin{itemize}
\item For a planar, vertically focusing undulator, any instability inside the undulator can be avoided with the exception of a single set of points, but only at the expense of an increased quad strength and an increasing variation of the $\beta$-function in the horizontal, plane.
\item Due to the phase advance variation as the undulator gap is closed, there are at regular distances points, where the quad setting has no influence on the stability. In these cases, the system is always unstable, unless the undulator is place in the center between the quads.
\item The shorter the intersection between undulators, the smaller the growth of the quad currents needed as the field strength of the undulator increases.
\item There are two kind of instabilities. One kind requires a FOFO-lattice in order to keep the optics stable and the instability becomes smaller as the focusing of the undulator increases, the other is caused by the defocusing at the end of the undulator and increases as the undulator focusing increases, given the correct phase advance. This latter instability can only  be compensated with increased focal strength by the quads in the undulator focusing plane, this drives the instability in the horizontal plane to the edge of instability.
\end{itemize}
Matching into the undulator has not been adjusted, but has no influence on the stability and little influence on the variation of the $\beta$-function along the undulator for the parameter sets studied at SXFEL. However, also focusing behind the undulator has not been adjusted. If undulators upstream are opened one-by-one, the procedure works fine. If they are opened starting at the end, the beam blows up in the part where the undulator has been opened in some cases. Even though the undulators are opened and therefore lasing is not influenced and probably also the opened undulators cannot be damaged, it has consequences for measurements in a diagnostics behind the undulator and the machine protection system could be triggered to stop the beam because of this. Further study is therefore definitely needed into optimization of the matching conditions.
\begin{figure}[!t]
	\centering
	 \includegraphics*[width=\columnwidth]{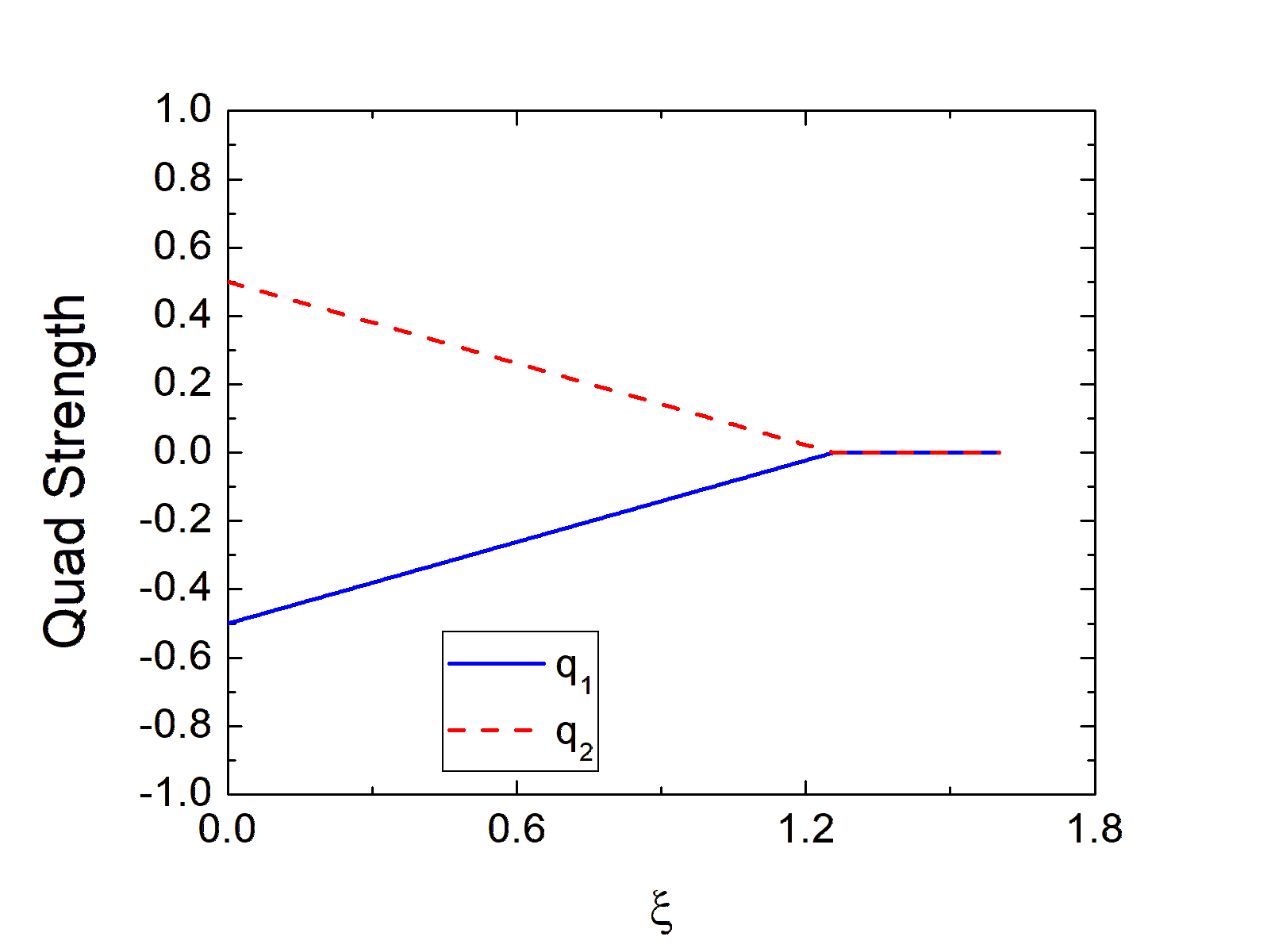}
	 \includegraphics*[width=\columnwidth]{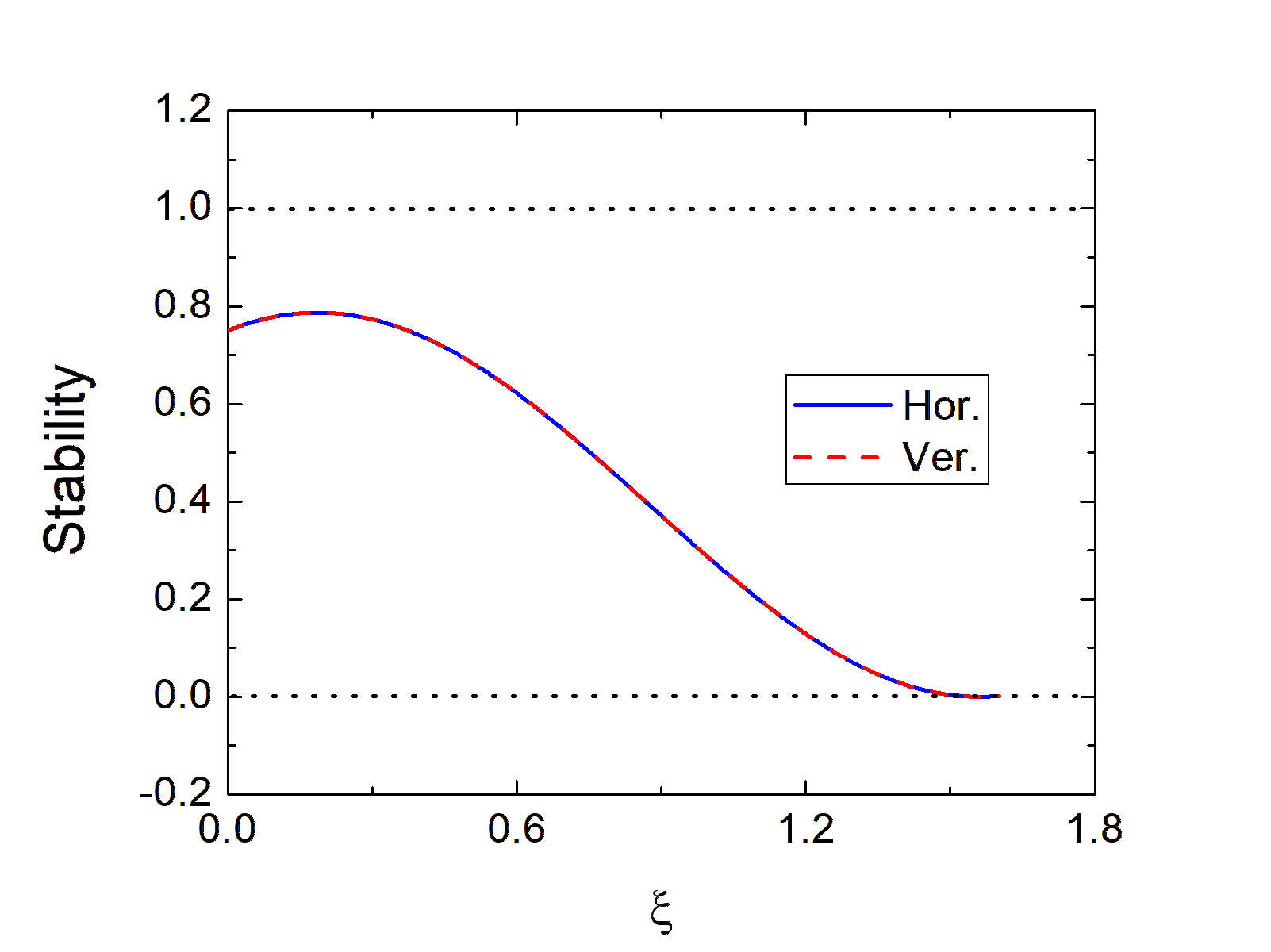}
	 	\caption{Quad settings (left) and stability function for the SXFEL SBP undulator, assuming a helical undulator with an energy down to 0.6~GeV. The stability is identical in both planes.}
	\label{fig:SXFEL-stability-helical}
\end{figure}

This paper has been completely dedicated to the case of a planar undulator. However, the undulator at the SBP-line at SXFEL has space reserved for devices with variable polarization and therefore can also be focusing in both planes. This means, the exchanging focusing between horizontal and vertical plane (FODO to FOFO) no longer works. With the undulator focusing in both planes, the quad focusing can only go to zero in the optimal case. Based on symmetry, it is also clear that for both planes to be stable, $q_1=-q_2$. Violating this condition can improve the situation in one plane, but always at the expense of the stability in the other plane. When the space is completely filled with undulators, the focusing starts with a FODO-structure while the undulators are open and are reduced equally to zero when the undulators are closed before the first instability occurs, and remain there. Before, the focusing was changed from a FODO to a FOFO focusing to keep the beam stable in both planes, but in this case, that is not needed. This will always give a stable optics for all values of $\xi$. However, if there are drift spaces between quads and undulators, as will always be the case, it is unclear if the strong defocusing by the undulator can be compensated in both planes simultaneously. Because $q_1=-q_2$, it may require even stronger focusing by the quads to move the focal point of quad and undulator combined in both planes to increase the stability. This results in very strong quad focusing and very large variation in $\beta$. Since this system has not been studied in a general context, no clear statement is possible. However, the first instability can be avoid by reducing the quad strength to zero, as was discussed in the case without drift spaces. In case of SXFEL, for example, the SBP-line can be stable down to an energy of 0.45~GeV, even for a helical undulator, for a moderate increase in quad strength by a factor of 3 compared to the nominal settings for a FODO structure. Considering the low beam energy, with the currents through the quads at very low value, this is well within specifications.

Therefore, for almost all machines, both helical and planar undulators can be used with the same automatic quad settings introduced at FLASH2 to keep the electron beam confined to a small size. This will reduce the risk of beam loss, make the machine more reproducible by avoiding manual change of the focusing and therefore make the operation more reliable for users.

\appendix

\section{Stability of the combined quad and undulator focusing}

We first define several parameters, as shown in Fig.~\ref{fig:Geometry}. We start with the components, with quadrupoles  $Q1$ and $Q2$ both approximated in the so-called thin-lens approximation, distance between the quads is $ l $ and an undulator with length $ l _u=a l $, where $0<a<1$, and a phase advance produced by the undulator of $\sqrt{a}\xi=K_{rms}k_u  l _u/\gamma$, where $2\pi\lambda_u=k_u$, $\gamma$ the Lorentz factor and $K_{rms}$ the rms value of the undulator strength. In matrix-form, the elements (quad, undulator and drift) can be described by

 \begin{equation*}
\begin{pmatrix}
1				 &0 \\
-\frac{q_i}{ l }				& 1	 \\
\end{pmatrix}\,,\
\end{equation*}
 \begin{equation*}
\begin{pmatrix}
\cos(\sqrt{a}\xi)				 &\frac{a l  \sin(\sqrt{a}\xi)}{\sqrt{a}\xi} \\
-\frac{\sqrt{a}\xi \sin(\sqrt{a}\xi)}{a l }			& 1\cos(\sqrt{a}\xi) \\
\end{pmatrix}\,,\
\end{equation*}
 \begin{equation*}
\begin{pmatrix}
1				 &(\frac{1}{2}\pm b)(1-a) l  \\
0				& 1	 \\
\end{pmatrix}
\end{equation*}
$q_i$ being the inverse focal length of the quadrupole $Q_i$, normalized to the distance between the quads, $-0.5<b<0.5$ is the position of the undulator between the quads, $b=0$ meaning a centered undulator.
A quadrupole focusing in one plane, is automatically defocusing in the other plane, which means it changes sign. In principle, a planar undulator has a slight defocusing effect in the other plane, which is neglected throughout this paper. In front and behind the undulator, there is a drift of length $(1/2\pm b)(1-a) l $.

The transfer map of half of the cell is a symplectic matrix:
\begin{equation}
M_{1/2} =
\begin{pmatrix}
B_{a,b}(\xi) - A_{a,b}(\xi) q_2 & A_{a,b}(\xi) l \\
M_{21} & B_{a,-b}(\xi) - A_{a,b}(\xi) q_1
\end{pmatrix}
\end{equation}
where the formulae of functions $A_{a,b}(\xi) $ and $B_{a,b}(\xi) $ are:
\begin{widetext}
\begin{align}
A_{a,b}(\xi) &= (1 - a)  \left[\,\cos(\sqrt{a}\xi) -  \frac{(1 - a)}{4 a} (1 -  4 b^2) \sqrt{a}\xi \sin(\sqrt{a}\xi)\right] +   a\, {\sin}c(\sqrt{a} \xi) \label{eq:A}\\
B_{a,b}\xi) &= \cos( \xi)  - \frac{(1 - a)}{2 a} (1 + 2 b)  \sqrt{a}\xi \sin(\sqrt{a} \xi) \label{eq:B}
\end{align}
\end{widetext}
It's only stable when $S_{a,b}(q_1,q_2,\xi)  \in [0,1]$, as discussed in Ref~\cite[p.~310]{Wiedemann2015}.

If $A_{a,b}(\xi)=0$,
\begin{equation}
M_{1/2} =
\begin{pmatrix}
B_{a,b}(\xi) & 0 \\
M_{21} & B_{a,-b}(\xi)
\end{pmatrix}
\end{equation}
Since $\det(M_{1/2}) \equiv 1$, that means $B_{a,-b}(\xi) = 1/B_{a,b}(\xi)$. The stability then becomes:
\begin{equation}
S = \frac14 \left( B_{a,b}(\xi) + \frac1{B_{a,b}(\xi)} \right)^2 \ge 1 \label{eq:S:A=0}
\end{equation}
So it's only stable when $B_{a,b}(\xi) = B_{a,-b}(\xi) = \pm 1$. This is true when $b=0$ because $\xi \notin \mathbb{N}$. Therefore (take $b=0$ in Eq.~\eqref{eq:A}):
\begin{equation}\label{eq:cos-sin:b=0}
\cos\xi = \frac12 \left(\frac{(1-a) \xi}{2a} - \frac{2a}{(1-a)\xi} \right) \sin\xi
\end{equation}
Square both sides:
\begin{equation}
\cos^2\xi = \frac14 \left(\frac{(1-a) \xi}{2a} - \frac{2a}{(1-a)\xi} \right)^2 \sin^2\xi = 1 - \sin^2\xi
\end{equation}
Combine the $\sin^2\xi$ terms:
\begin{equation}\label{eq:sin2:b=0}
\frac14 \left(\frac{(1-a) \xi}{2a} + \frac{2a}{(1-a)\xi} \right)^2 \sin^2\xi = 1
\end{equation}
Now look at Eq.~\eqref{eq:B} and take $b=0$:
\begin{equation}
\begin{split}
B_{a,b}(\xi) &= \cos\xi - \frac{1-a}{2a} \xi \sin\xi \\
\intertext{(apply Eq.~\eqref{eq:cos-sin:b=0})}
&= \frac12 \left(\frac{(1-a) \xi}{2a} - \frac{2a}{(1-a)\xi} \right) \sin\xi - \frac{1-a}{2a} \xi \sin\xi \\
&= - \frac12 \left(\frac{(1-a) \xi}{2a} + \frac{2a}{(1-a)\xi} \right) \sin\xi \\
\intertext{(apply Eq.~\eqref{eq:sin2:b=0})}
&= \pm1
\end{split}
\end{equation}
Therefore, when $A_{a,b}(\xi)=0$, $B_{a,b}^2(\xi)$ always equals to $1$ if we take $b=0$, meaning it's always stable at $A_{a,b}(\xi)=0$.

\section*{Acknowledgement}
We wish to thank Stephan Tzenov for discussions and all colleagues from SXFEL for their support. Bart Faatz was funded by the Chinese Academy of Sciences President’s International Fellowship Initiative (PIFI), Grant No. 2020FSM0003.

\end{document}